\pdfoutput=1
\documentclass[reprint,aip,jcp,showpacs,twocolumn]{revtex4-1}
\usepackage{mathptmx}
\usepackage{epsfig,amsopn}
\usepackage{graphicx}
\usepackage{amsmath,amssymb}
\usepackage{natbib}
\usepackage{xcolor}
\usepackage{braket}
\usepackage{float}
\usepackage{enumerate}
\usepackage{mathtools}
\usepackage[symbol]{footmisc}
\usepackage{appendix}
\usepackage{stackengine}
\allowdisplaybreaks[1]
\usepackage{hyperref}
\usepackage{ulem}

\newcommand{\eref}[1]{Eq.~(\ref{#1})}%
\newcommand{\fref}[1]{Fig.~\ref{#1}} %
\newcommand{\aref}[1]{Appendix~\ref{#1}}%

\def\bea{\begin{eqnarray}}
\def\eea{\end{eqnarray}}
\def\bal#1\eal{\begin{align}#1\end{align}}

\hypersetup{
    colorlinks=true,
    linkcolor=blue,
    citecolor=blue, 
   urlcolor=blue,
    }
\begin{document}

\title{Rate enhancement of gated drift-diffusion process by optimal resetting}

\author{Arup Biswas}
\author{Arnab Pal}
\email{arnabpal@imsc.res.in}
\affiliation{\noindent \textit{The Institute of Mathematical Sciences, CIT Campus, Taramani, Chennai 600113, India \& 
Homi Bhabha National Institute, Training School Complex, Anushakti Nagar, Mumbai 400094,
India}}
\author{Debasish Mondal}
\affiliation{\noindent \textit{Department of Chemistry, Indian Institute of Technology Tirupati, Tirupati 517619, India}}
\author{Somrita Ray}
\email{sray@ed.ac.uk}
\affiliation{\noindent \textit{Department of Chemistry, Indian Institute of Technology Tirupati, Tirupati 517619, India}}
\affiliation{\noindent \textit{SUPA, School of Physics \& Astronomy, University of Edinburgh,
Peter Guthrie Tait Road,
Edinburgh EH9 3FD, United Kingdom. }}
\date{\today}
\begin{abstract}
`Gating' is a widely observed phenomenon in biochemistry that describes the transition between the activated (or open) and deactivated (or closed) states of an ion-channel, which makes transport through that channel highly selective. In general, gating is a mechanism that imposes an additional restriction on a transport, as the process ends only when the `gate' is open and continues otherwise. When diffusion occurs in presence of a constant bias to a {\it gated} target, i.e., to a target that switches between an open and a closed state, the dynamics essentially slows down compared to {\it ungated} drift-diffusion, resulting in an increase in the mean completion time, $\langle T^G\rangle>\langle T\rangle$, where $T$ denotes the random time of transport and $G$ indicates gating. In this work, we utilize stochastic resetting as an external protocol to counterbalance the delay due to gating. We consider a particle in the positive semi-infinite space that undergoes drift-diffusion in the presence of a stochastically gated target at the origin and is moreover subjected to a rate-limiting resetting dynamics. Calculating the minimal mean completion time $\langle T_{r^{\star}}^G\rangle$ rendered by an optimal resetting rate $r^{\star}$ for this exactly-solvable system, we construct a phase diagram that owns three distinct phases: (i) where resetting can make gated drift-diffusion faster even compared to the original ungated process, $\langle T^G_{r^{\star}}\rangle<\langle T\rangle<\langle T^G\rangle$, (ii) where resetting still expedites gated drift-diffusion, but not beyond the original ungated process, $\langle T\rangle \leq \langle T^G_{r^{\star}}\rangle <\langle T^G\rangle$, and (iii) where resetting fails to expedite gated drift-diffusion, $\langle T\rangle<\langle T^G\rangle \leq \langle T^G_{r^{\star}}\rangle$. We also highlight various non-trivial behaviors of the completion time as the resetting rate, gating parameters and the geometry of the set-up are carefully ramified. Gated drift-diffusion aptly models various stochastic processes such as chemical reactions that exclusively take place for certain activated state of the reactants. Our work predicts the conditions where stochastic resetting can act as a useful strategy to enhance the rate of such processes without compromising on their selectivity.

 \end{abstract}

\maketitle
\section{Introduction}
\label{intro}
\begin{figure}[t]
    \centering
    \includegraphics[width=8.7cm]{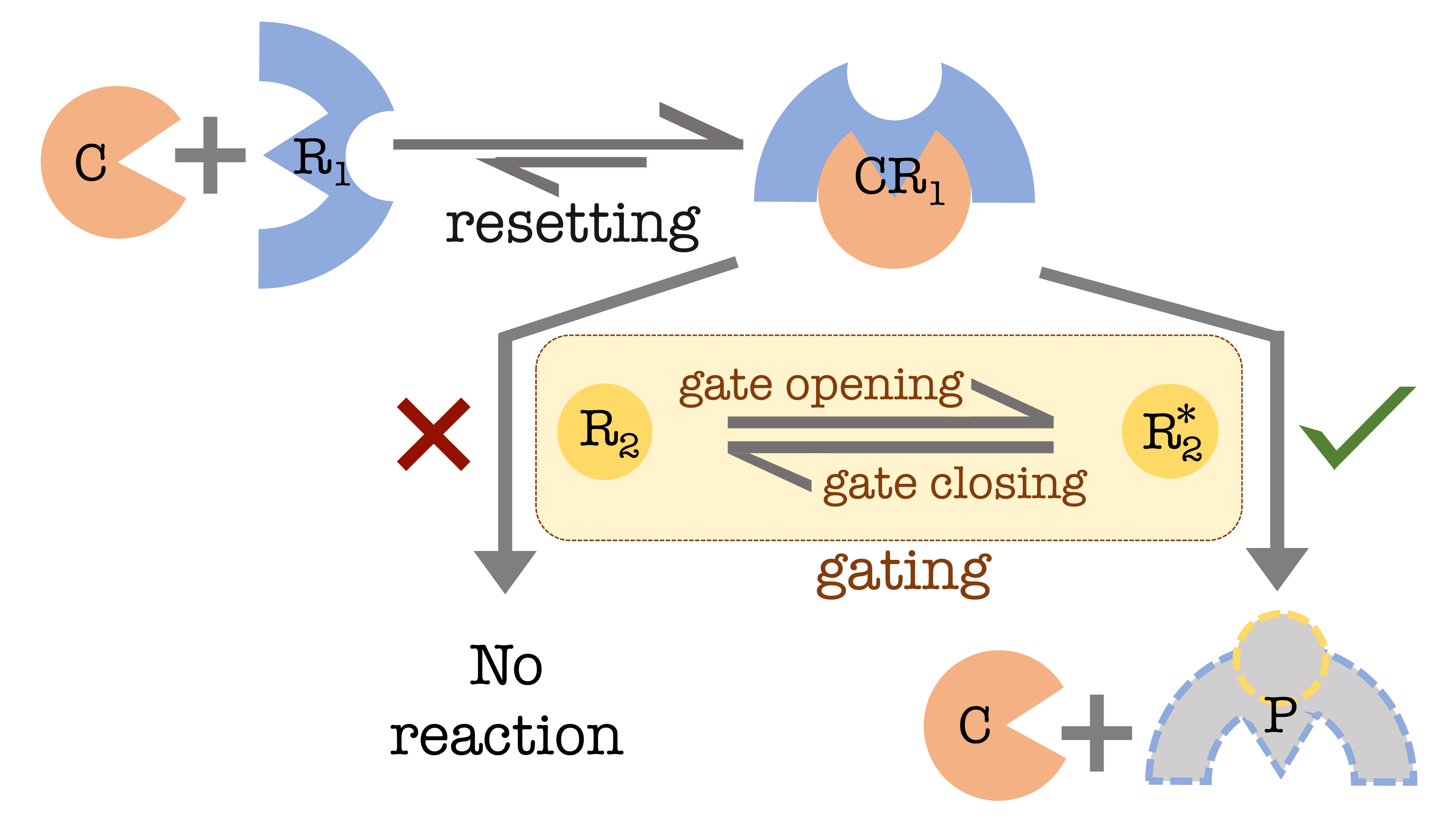}
    \caption{Scheme for a gated chemical reaction between two reactants $R_1$ and $R_2$, catalysed by $C$. In the first step, the catalyst initiates the reaction by binding reversibly with $R_1$ to generate $CR_1$, a metastable intermediate: $C+R_1\stackrel{}{\rightleftharpoons}CR_1$. In the second step, $CR_1$ reacts selectively with  $R^{\star}_2$ (the {\it reactive} or {\it open-gate} state of $R_2$) to generate the product $P$ and liberate the catalyst: $CR_1+R_2^{\star}\to C+P$. This step can be modeled by gated drift-diffusion, while the unbinding of $C$ from $R_1$ is essentially resetting. }
    \label{Fig1}
\end{figure}

`Gating' in biochemistry typically refers to the transition between the {\it open} and {\it closed} states of an ion channel; the ions are allowed to flow through the channel only when it is open\cite{ic1}. In a gated chemical reaction, the reactant molecules switch between a reactive and a non-reactive state; the collisions between the reactants {\it must} happen in their reactive states for a successful reaction. Thus, gating is a signature of a constrained reaction/transport, be it an enzyme finding the correct substrate or a protein finding the target site along a DNA strand or associating to a cleavage site on a peptide. Given their very generic nature, it is no wonder why gated processes showcase a myriad of applications spanning across fields from chemistry \cite{gating-review,gated-6,gated-8,gated-9}, physics \cite{gated-10,gated-11,gated-12,gated-13,gated-14,gated-15,gated-16} to biology \cite{gated-bio-1,gated-bio-2,gated-bio-3}.\\
\indent
Since the seminal works of Szabo \textit{et al} \cite{gated-8,gated-10,Szabo-*}, gated processes have 
gained considerable attention across a
wide panorama of applications such as 1D gated continuous-time
and discrete-space random search process in confinement \cite{gated-12}, intermittent switching dynamics of a protein undertaking facilitated diffusion on a DNA strand \cite{conform-1}, 3D gated diffusive search process with different diffusivities \cite{gated-13}, gated active particles  \cite{gated-active}, to name a few. Gopich and Szabo explored the possibility of multiple gated particles/targets in a model with intrinsic reversible binding \cite{gated-15}. More on a mathematical side, Lawley and Keener established a connection between a radiative/reactive boundary and a gated boundary in diffusion controlled reactions \cite{robin}. There has been a renewed interest in the field emanating from the works by Mercado-Vásquez and Boyer on a 1D gated diffusive process on the infinite line \cite{gated-0}, Scher and Reuveni on gated reactions on arbitrary networks including random walk models both in continuous \cite{gated-2} and discrete time \cite{gated-3} and Kumar \textit{et al} on threshold crossing events of a gated process \cite{gated-4} and inference of first-passage times from the detection times of gated diffusive first-passage processes \cite{inference-gated}. In a similar vein, in this work, we delve deeper into a gated diffusive process and in particular, focus on to design principles to improve efficiency of a gated reaction [see Fig. (\ref{Fig1})].\\
\indent
For a gated diffusive process to be complete, a certain condition that mimics the open-gate-scenario, imposed either on the diffusing entity or on the target that it diffuses to, needs to be fulfilled. This additional restriction imposed due to gating certainly makes the process more selective, which is essential for the associated biochemical system to function properly. The cost for this selectivity, however, reflects on the completion time, which makes a gated diffusive process essentially slower than the corresponding ungated one. However, nature has its own way to curtail such situation to allow effective reactions. For example, consider a chemical reaction such as in Fig. (\ref{Fig1}) where an unbinding or a resetting from a metastable state can lead to a facilitated reaction. The effects of such resetting events are proven to be crucial not only in such chemical reactions \cite{chemical-1-r,chemical-2-r}, but also in the backtracking of RNA polymerase \cite{RNA-1-r} and disassociation kinetics of RhoA in the membrane \cite{Rhoa-1-r}. Motion of the reactants or the morphogens which get produced constantly from a certain place inside the cell before degrading in time (due to their finite lifetime) can also be interpreted as stochastically restarted processes \cite{morpho}. More on the physics side, it has been established in the pioneering work of Evans and Majumdar \cite{Restart1} that stochastic resetting can be utilized as a powerful strategy to expedite the completion time of a 1D diffusive search process. This remarkable result led to many fascinating works where it was shown that indeed this intermittent resetting strategy can benefit the search processes conducted by diffusion controlled \cite{PalJphysA,FP-pot-0,FP-pot,FP-interval,JCP-1,JCP-2,JCP-3,DM_JPhysA,RayPRE,RW-OLB} and non-diffusive stochastic processes \cite{CB-2,Restart7,Restart11,records} [see here \cite{Evansrev2020} for an extensive review of the subject]. Single particle experiments using optical traps have also paved the way of our understanding of resetting modulated search processes \cite{expt-1,expt-2}. Albeit these advances, there has been a persistent void in the understanding of resetting induced gated processes until recently when resetting mechanism has been used to reduce the average completion time of a gated diffusive process in 1D \cite{gated-1,gated-5}. \\
\indent
While resetting is a useful strategy to benefit \textit{diffusive} transport of ions/ molecules in unbounded phase space, the same can not be said for a \textit{drift-diffusive} search process. There, resetting can be useful only if the rate of diffusive transport is higher compared to the rate of driven transport \cite{PECLET,branching,channel,local}. However, if the drift supersedes the transport of molecules across the channel, resetting can only hinder the completion rendering a higher transit time. This crucial interplay can then be understood in terms of the so-called P\'eclet number, which is a ratio between the diffusion- and drift-dominated microscopic time-scales \cite{PECLET,branching}. In fact, the role of resetting can be quantified within a universal set-up of first-passage under restart which essentially teaches us that resetting will be always beneficial if the underlying process without resetting has a coefficient of variation (often known as the signal to noise ratio) -- a statistical measure of dispersion in the random completion time, defined as the ratio of its standard deviation to its mean --  greater than unity \cite{inspection,Restart7,Restart8}. These observations naturally set the stage to understand the role of resetting in a gated drift-diffusive search process which, to the best of our knowledge, has not been studied so far. In particular, the central objective of this work is to understand whether resetting can enhance the completion rate of a gated drift-diffusive search process and if so, under what conditions. Unraveling the intricate role of drift, diffusion and gating along with that of resetting will be at the heart of this study.\\
\indent
To illustrate the set-up of the present work, let us consider a diffusive transport process confined to the positive semi-infinite space in presence of a constant drift which is directed towards a target that randomly switches between an open and a closed state with constant rates. This is a general scenario mimicked by, e.g., a chemical reaction [see \fref{Fig1}], where the collisions between reactants [$CR_1$ and $R_2$] lead to the formation of product only when at least one of the reactants is in an activated state [when $R_2$ exists as $R_2^{\star}$] and not otherwise. Since resetting is an integral part of any consecutive chemical reaction that has a reversible first-step [when $C$ binds to, or unbinds from, $R_1$], the reaction scheme shown in \fref{Fig1} can be modeled as gated drift-diffusion with resetting. In this paper, we study the completion time statistics for this general set-up. In particular, we perform a comprehensive analysis to understand the role of optimal resetting in enhancing the rate of such transport and construct a complete phase diagram that distinguishes between the phases where (i) resetting accelerates gated drift-diffusion beyond the original (without resetting) ungated process, (ii) resetting still proves itself to be beneficial by improving the rate of gated drift-diffusion, but not beyond the original ungated process and (iii) resetting can not expedite gated drift-diffusion.\\
\indent
The rest of the paper is organized as follows. In Section \ref{completion-time}, we describe the model and follow a Fokker-Planck approach to write down the governing equations for gated drift-diffusion process that is subject to resetting. In the same Section, we solve those equations to calculate the mean completion time, denoted $\langle T_r^{G}\rangle$. We calculate the optimal resetting rate $r^{\star}$ that minimizes $\langle T_r^{G}\rangle$, and thereby investigate the conditions for resetting to expedite the process completion in Section \ref{restart-criterion}. There, various limits of the system parameters are also examined in detail. We calculate the maximal speedup that can be achieved, within our set-up, by the optimal resetting rate in Section \ref{speed-up}. In Section \ref{full-phase}, we construct a full phase diagram that characterizes all the possible regimes underpinning the role of resetting in the process completion. We conclude with a brief summary and outlook of our work in Section \ref{summary}. Some of the detailed derivations have been moved to the Appendix for brevity. 
\vspace{-0.2cm}
\section{Completion time statistics}
\label{completion-time}
\begin{figure}[t]
    \centering
    \includegraphics[width=8.5cm]{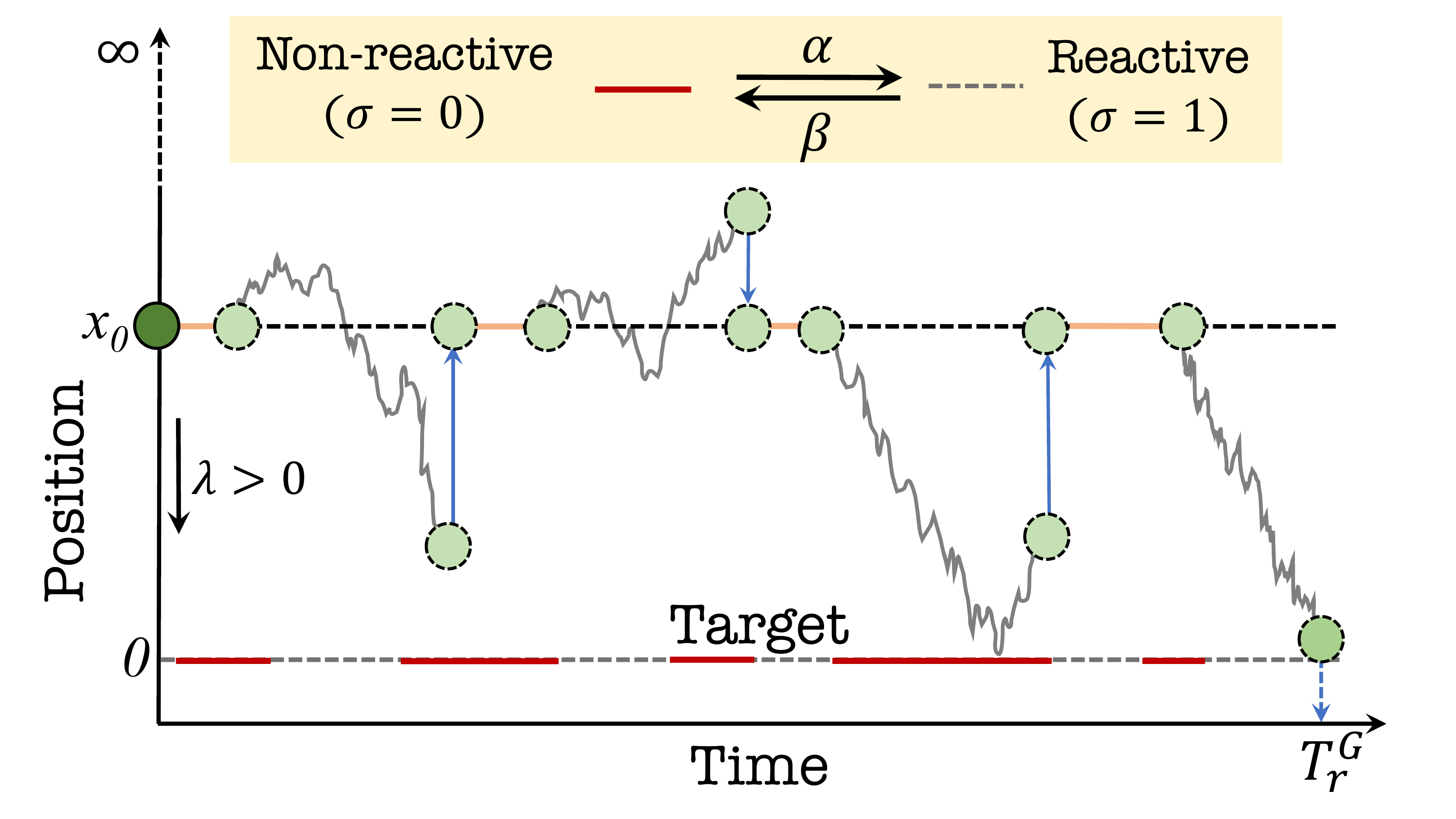}
    \caption{Schematic diagram of a gated drift-diffusion process in semi-infinite space with resetting, where the target (placed at the origin) switches stochastically between a reactive state ($\sigma=1$) and a non-reactive state ($\sigma=0$). The transition from the non-reactive to the reactive state takes place with a constant rate $\alpha>0$, and the opposite transition takes place with a constant rate $\beta>0$. When the particle, starting at $x_0$, hits the target in its reactive state (dashed line), the process ends. $T_r^G$ marks the random completion time of the gated drift-diffusion process with resetting.}
    \label{Fig2}
\end{figure}

We start by casting the problem of gated chemical reaction [$CR_1+R_2^{\star}\to C+P$, introduced in \fref{Fig1}] as gated drift-diffusion. A convenient way to do so is to map the reaction coordinate associated to the reactants $CR_1+R_2^{\star}$ onto the starting position [$x_0>0$, see \fref{Fig2}] of a particle that undergoes Brownian motion. Similarly, the reaction coordinate for the products $C+P$ can be mapped onto the position of a target [placed at the origin, see \fref{Fig2}]. The chemical potential drive, which governs the reaction to its completion, can then be translated to a constant bias $\lambda$ that the particle experiences while it diffuses to the target with a diffusion coefficient $D$. Note that $\lambda$ is considered to be positive when it acts towards the target and negative when it acts away from the target. Such effective one-dimensional projection of the energy landscape of an enzyme or protein conformation is a well-adapted approach in the literature -- see \cite{effect-pot-1,effect-pot-2,effect-pot-3} for more details. Nonetheless, there are a few key assumptions that we make during the mapping. First, we take into account an well-established fact that the chemical master equations for the discrete states can be coarse-grained into the Fokker-Planck equations for the reaction coordinates in the continuous space using a system size expansion\cite{kampen,gardinar}. However, this conversion generically renders the drift and diffusion terms to be spatially dependent. In other words, the reaction coordinate associated to the catalysis process $CR_1+R_2^{\star}\to C+P$ is expected to diffuse in an {\it arbitrary} energy landscape, where the product state $C+P$ is usually denoted by the global minimum. However, to simplify the problem, we replace the general potential by a linear one, which leads to a constant drift velocity $\lambda$ towards (or away from) the gated target. This is the second assumption that lies behind our analysis. Albeit these simplified approximations, the major advantage here is the elegant analytical tractability of the model from which one can also unveil the crucial interplay between the gated boundary and resetting/unbinding mechanism.\\
\indent
To model gating, the target is considered to randomly switch between a reactive ($\sigma=1$ state in \fref{Fig2} that resembles $R_2^{\star}$ in \fref{Fig1}) and a non-reactive state [$\sigma=0$ state in \fref{Fig2} that resembles $R_2$ in \fref{Fig1}]. The conversion from the non-reactive to the reactive state happens with a constant rate $\alpha>0$ and that from the reactive to the non-reactive state happens with a constant rate $\beta>0$ [\fref{Fig2}]. Therefore, the {\it reactive occupancy} of the target, i.e., the probability of finding the target in its reactive state is $p_r\coloneqq\alpha/(\alpha+\beta)$. Similarly, the {\it non-reactive occupancy} or the probability that the target is in its non-reactive state is $p_{nr}\coloneqq(1-p_r)=\beta/(\alpha+\beta)$. In analogy to the chemical reaction, the process ends only when the particle hits the target in its reactive state. When it hits the target in its non-reactive state, it simply gets reflected and continues to diffuse.\\
\indent
As mentioned in the Introduction, the unbinding of the catalyst $C$ from $R_1$ can be interpreted as resetting at $x_0$. For simplicity, here we consider Poissonian resetting with a constant rate $r$. In the gated drift-diffusion scenario, this means that the particle is taken back to $x_0$ after stochastic intervals of time, taken from an exponential waiting time distribution with mean $r^{-1}$. Note that here we assume that the resetting is instantaneous and once the particle is reset at $x_0$, it immediately starts moving again. In other words, we neglect any refractory period, i.e., idle time after each resetting event before drift-diffusion resumes, considering that the time-scale of waiting at $x_0$ after reset [which maps to the time required for $C$ to bind $R_1$ in \fref{Fig1}] is much smaller compared to that of either drift-diffusion or resetting. We also assume that the intermittent dynamics of the target are independent of
resetting similar to \cite{gated-1}.

The fundamental quantity of interest here is the random completion time of the gated chemical reaction, i.e., the {\it first-passage time} \cite{REDNER} of the diffusing particle from $x_0$ to the target placed at the origin. To calculate the mean first-passage time (MFPT), we first need to write the backward Fokker-Planck equation for the survival probability of this system, which is the total probability to find the particle in the interval $[0,\infty)$ at a time $t$, provided the initial position is $x_0$. We denote $Q_{\sigma}(t|x_0)$ as the joint survival probability, i.e, the probability that the particle has not been absorbed by the target up to time $t$, given the initial position $x_0$ and initial target state $\sigma(t=0)$. One can then construct the backward Fokker-Planck equations \cite{gated-1} in terms of the initial position $x_0$, which now serves as a variable 
\begin{align}
\frac{\partial Q_{0}(t|x_0)}{\partial t}=&-\lambda \frac{\partial Q_{0}(t|x_0)}{\partial x_0}+D\frac{\partial^2 Q_{0}(t|x_0)}{\partial x_0^2}+\alpha [Q_{1}(t|x_0)\nonumber\\
&-Q_{0}(t|x_0)]+r[Q_{0}(t|x_r)-Q_{0}(t|x_0)],\nonumber\\
\frac{\partial Q_{1}(t|x_0)}{\partial t}=&-\lambda \frac{\partial Q_{1}(t|x_0)}{\partial x_0}+D\frac{\partial^2 Q_{1}(t|x_0)}{\partial x_0^2}+\beta [Q_{0}(t|x_0)\nonumber\\
&-Q_{1}(t|x_0)]+r[Q_{1}(t|x_r)-Q_{1}(t|x_0)].
    \label{bfpe}
\end{align}
Note that $x_r$ in the above set of equations indicates the resetting position which has to be distinguished from the variable $x_0$ initially and only at the end, has to be set $x_r=x_0$ self-consistently. The initial conditions for \eref{bfpe} are $Q_{\sigma}(0|x_0)=1$ and the boundary conditions are $Q_{1}(t|0)=0$ and $[\partial Q_{0}(t|x_0)/ \partial x_0]_{x_0=0}=0$, respectively. This indicates that the particle is absorbed at the target when the latter is reactive ($\sigma=1$) and is reflected from the target when it is non-reactive ($\sigma=0$) \cite{gated-1}. The {\it average} survival probability for the gated process can then be written by taking contributions from both the possibilities
\begin{equation}
Q_{r}^G(t|x_0)=p_r\;Q_1(t|x_0)
+(1-p_r) Q_0(t|x_0).
\label{qavg}
\end{equation}
Subsequently, we use subscript $r$ and superscript $G$ to indicate resetting and gating respectively in \eref{qavg} and rest of the paper. The Laplace transformation of \eref{qavg} gives
 \bal
 \tilde{Q}_{r}^G(s|x_0)=p_r\;\tilde{Q_1}(s|x_0)
+(1-p_r) \tilde{Q_0}(s|x_0),
\label{qavg_lt}
\eal
where $\tilde{Q}_{r}^G(s|x_0)\coloneqq\int_0^{\infty}dt\;e^{-st}Q_{r}^G(t|x_0)$ is the Laplace transform of $Q_{r}^G(t|x_0)$ and $\tilde{Q}_{\sigma}(s|x_0)\coloneqq\int_0^{\infty}dt\;e^{-st}Q_{\sigma}(t|x_0)$ are Laplace transforms of $Q_{\sigma}(t|y)$, respectively. The average MFPT, our observable of interest in this work, is also defined over the two random possibilities and thus reads
 \bal
\langle T_{r}^G(x_0)\rangle = p_r\langle T_{1}(x_0)\rangle+
 (1-p_r)\langle T_{0}(x_0)\rangle,
\label{tavg_def}
\eal
where $\langle T_{\sigma}(x_0)\rangle$ is the MFPT when the initial state of the target is $\sigma$, given by $\langle T_{\sigma}(x_0)\rangle=\int_0^{\infty}\;dt Q_{\sigma}(t|x_0)=\tilde{Q}_{\sigma}(s|x_0)|_{s=0}$, since $-\partial Q_{\sigma}(t|x_0)/\partial t$ is the associated first-passage time distribution \cite{REDNER,gardinar}. Therefore, the average MFPT reads
 \bal
 \langle T_{r}^G(x_0)\rangle 
 =\int_0^{\infty}dt\;Q_{r}^G(t|x_0) =\tilde{Q}_{r}^G(s|x_0)|_{s=0}.
\label{tavg_q}
\eal
For the rest of the paper, we will write $\langle T_{r}^G\rangle$ instead of $\langle T_{r}^G(x_0)\rangle$ for brevity.\\
\indent 
Solving \eref{bfpe} in the Laplace space, plugging in the solutions, i.e., $\tilde{Q}_{\sigma}(s|x_0)$s, into \eref{qavg_lt} to calculate $\tilde{Q}_{r}^G(s|x_0)$ and finally setting $s=0$ in the resulting expression of $\tilde{Q}_{r}^G(s|x_0)$ following \eref{tavg_q}, we obtain [see \aref{appa} for detailed derivation] the explicit expression of the average MFPT that reads
\bal
\langle T_{r}^G\rangle=\frac{1}{r}\left(e^{\mu_1 x_0}-1\right)+
\frac{\beta \mu_1}{\alpha \mu_2} \left(\frac{1}{r} + \frac{e^{-
   \mu_2 x_0}}{\alpha + \beta}\right)e^{\mu_1 x_0},
\label{mfpt_av}
\eal
where $\mu_{1}=(-\lambda +\sqrt{\lambda ^2+4 D r})/2D>0$ and $\mu_{2}=(-\lambda +\sqrt{\lambda ^2+4 D (\alpha +\beta +r)})/2D>0$. Therefore, for pure diffusion with gating, i.e., when $\lambda\to 0$, $\mu_1=\sqrt{r/D}$ and $\mu_2=\sqrt{(r+\alpha+\beta)/D}$ and \eref{mfpt_av} boils down to \cite{gated-1} 
\begin{align}
\langle T_{r}^G\rangle=\frac{e^{\sqrt{\frac{r}{D}}x_0}-1}{r}
+
\frac{\beta e^{\sqrt{\frac{r}{D}}x_0}}{\alpha \sqrt{r[r+\alpha+\beta]}}\mbox{\hspace{-0.15cm}}\left[1+\frac{r e^{-\sqrt{\frac{r+\alpha+\beta}{D}}x_0}}{(\alpha+\beta)}\right].
\label{mfpt_av_d}
\end{align}
Moreover, in the absence of gating, i.e., when $\beta\to 0$, \eref{mfpt_av} reduces to $\langle T_{r}\rangle=[\exp{({x_0(\sqrt{\lambda^2+4Dr}-\lambda)/2D}})-1]/r$, which is the exact expression for the MFPT for ungated drift-diffusion with Poissonian resetting \cite{PECLET}. \\
\begin{figure}[t!]
    \centering
     \includegraphics[width=8.1cm]{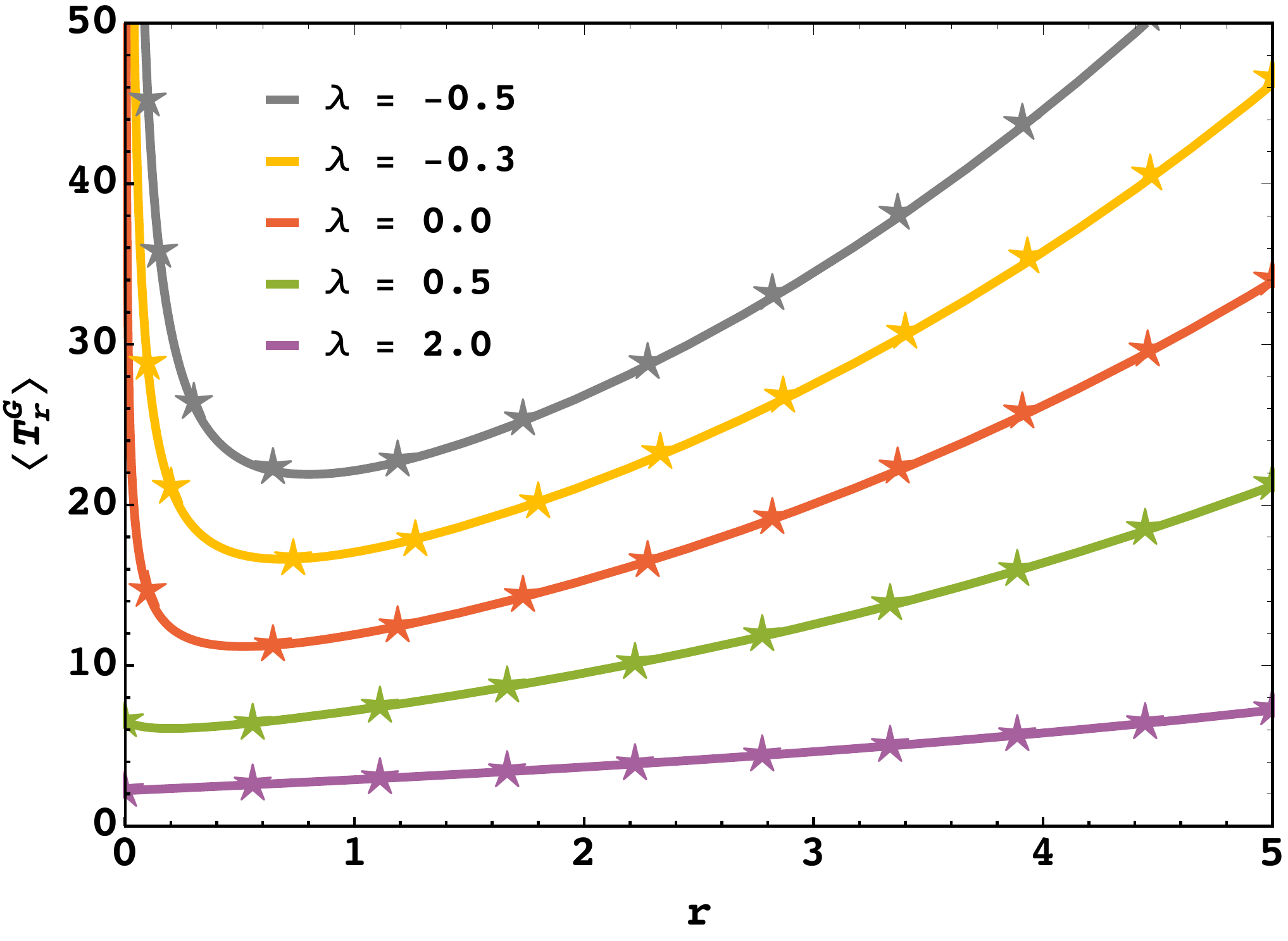}
      \caption{The average MFPT, $\langle T_{r}^G\rangle$, vs the resetting rate $r$ for different values of the drift velocity $\lambda$. The lines represent analytical results following \eref{mfpt_av} and the symbols represent results from numerical simulations [see \aref{appd} for details]. The curves with $\lambda>0$ denote cases where the drift acts towards the gated target and that with $\lambda<0$ denote otherwise. The variation of $\langle T_{r}^G\rangle$ with $r$ is always non-monotonic for $\lambda\leq 0$, however, for $\lambda>0$ it is non-monotonic for lower values of $\lambda$, but monotonic for sufficiently high values of $\lambda$. Here we take $x_0=2$, $D=1$, $\alpha=0.5$, and $\beta=0.5$ for all cases.   }
    \label{Fig3}
\end{figure}
\indent
In \fref{Fig3}, we plot $\langle T_{r}^G\rangle$ as a function of the resetting rate $r$ for different values of the drift velocity $\lambda$, where the reactive occupancy of the target, $p_r$, is kept constant. It is evident from \fref{Fig3} that when $\lambda>0$, i.e., when the drift acts towards the target, the average MFPT shows a non-monotonic variation with $r$ for lower values of $\lambda$, indicating that the introduction of resetting expedites first-passage when the process is diffusion-controlled. In contrast, $\langle T_{r}^G\rangle$ increases monotonically with $r$ for sufficiently higher values of $\lambda$, meaning that the introduction of resetting delays first-passage when the process is drift-controlled. This trend can be explained in the following way. When diffusion dominates over the drift, the particle tends to diffuse away from the target. In such cases, resetting can effectively truncate those long trajectories, which reduces the overall first-passage time. In contrast, when the drift dominates over diffusion, the particle tends to execute a directed motion towards the target ($\lambda>0$); resetting can only hinder such transport resulting in a longer completion time. Note that resetting can accelerate first-passage even when the dynamics is drift-controlled, if the drive is away from the target ($\lambda<0$).\\
\indent
Summarizing, we see that either for pure diffusion or when the drift acts away from the target ($\lambda\leq 0$), the average MFPT consistently shows a non-monotonic variation with $r$. This means that the introduction of resetting always accelerates first-passage for $\lambda\leq 0$. Thus, a hallmark of {\it resetting transition} is apparent for $\lambda>0$, while no such transition is expected for $\lambda\leq 0$. Indeed, similar to ungated diffusive processes under resetting \cite{Restart1,PalJphysA}, introduction of resetting was shown to be always beneficial for pure diffusive gated process ($\lambda=0$) and no such transition was observed there \cite{gated-1}. Here, we focus on the scenario when the drift acts towards the target and then reveal the physical conditions under which resetting strategy turns out to be beneficial. In passing, we will also briefly discuss a situation where the system is confined between two boundaries (one additional reflecting boundary apart from the stochastically gated target). In analogy to a chemical reaction, the reflecting boundary is often used to mimic high activation energy barriers in reaction coordinate. We refer to \aref{appb} for this discussion, where we investigate the role of resetting in details.
\section{The optimal resetting rate and a phase diagram for expedited completion}
\label{restart-criterion}
The resetting transition is traditionally captured by the optimal resetting rate (ORR, denoted $r^{\star}$), defined as the rate of resetting that minimizes the average MFPT. 
\fref{Fig3} reveals that for pure diffusion ($\lambda=0$) the optimal resetting rate has a certain positive value (denoted $r^{\star}=r_0^{\star}$, not marked in \fref{Fig3}). With increase in the drift towards the target, ORR gradually decreases to finally become zero at a critical value of $\lambda$ (denoted $\lambda_c$, not shown in \fref{Fig3}), which marks the point of resetting transition. If we continue to increase $\lambda$ beyond $\lambda_c$, ORR remains zero. Note that  when the drift acts away from the target ($\lambda<0$), $r^{\star}$ increases as that drift becomes stronger. The optimal resetting rate $r^{\star}$ thus acts as an order parameter, in the similar spirit as in classical phase transition,  to explore the resetting transition. 
Since ORR minimizes $\langle T_{r}^G\rangle$, it can be calculated from the relation $d\langle T_{r}^G\rangle/dr|_{r=r^{\star}}=0$, which leads to a complicated transcendental equation that can not be solved analytically. 
\begin{figure}[t!]
    \begin{center}
     \includegraphics[width=8.2cm]{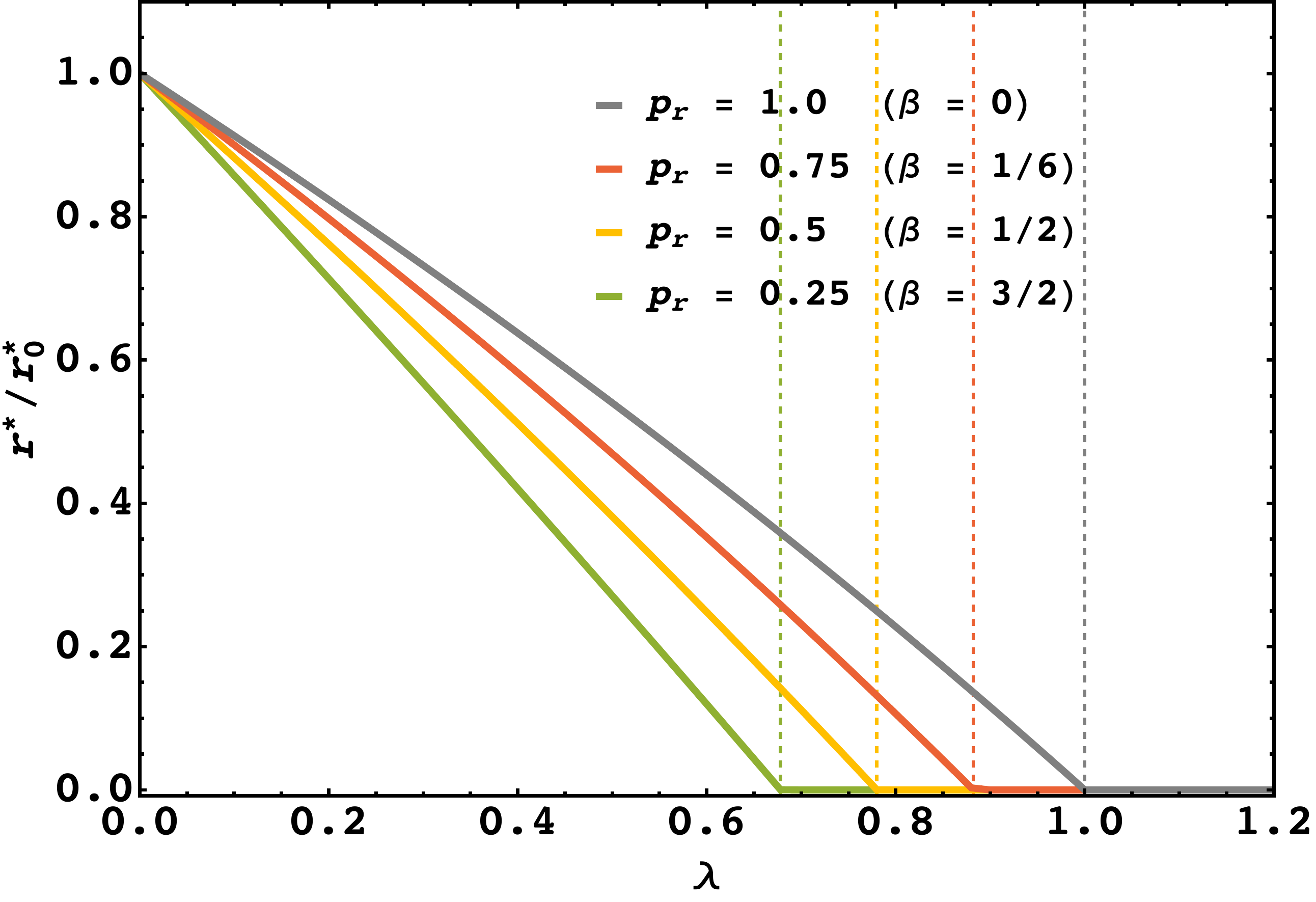}
    \end{center}
    \caption{The scaled optimal resetting rate, $r^{\star}/r^{\star}_0$, vs $\lambda$ for different values of $p_r$. Resetting transition in each case is observed at $\lambda=\lambda_c$, where the scaled ORR becomes zero, marked by dashed lines of the same color as the curve. For $\lambda<\lambda_c$, resetting expedites transport, whereas for $\lambda\geq\lambda_c$ it can not. Here we take $D=1$, $x_0=2$, and $\alpha=0.5$ [which leads to $p_r=1/(2\beta+1)$] for all cases. For ungated drift-diffusion [given by $p_r=1$, since $\beta\to0$], the resetting transition is observed at $\lambda_c=1$ (gray curve). For gated drift-diffusion with $p_r<1$, $\lambda_c$ decreases below unity.
      }
      \label{Fig4}
\end{figure}
Solving the same numerically, we calculate $r^{\star}$, the optimal resetting rate for $\lambda>0$. In a similar manner, the optimal resetting rate for pure diffusion (for $\lambda=0$, denoted $r^{\star}_0$) is also obtained in order to calculate the scaled ORR, $r^{\star}/r^{\star}_0$. \\
\indent
In \fref{Fig4}, we plot the scaled ORR with respect to $\lambda$ for different values of the reactive occupancy $p_r$ (we keep $\alpha$ constant and tune $p_r$ by solely changing $\beta$). The non-zero values of the scaled ORR for $\lambda<\lambda_c$ in \fref{Fig4} shows that resetting accelerates first-passage in that regime. In stark contrast, for $\lambda\geq\lambda_c$ the scaled ORR is zero, which shows that resetting can not accelerate first-passage in that regime. For the ungated ($p_r=1$) process, $\lambda_c=1$ for our choice of parameters, which is in exact agreement with earlier literature \cite{PECLET}. It is evident from \fref{Fig4} that $\lambda_c$ decreases when $p_r$ is decreased below unity, which means that resetting expedites first-passage time upto a critical drift (which is essentially smaller than $\lambda_c$ for the ungated process) when gating is introduced to the system. \\
\begin{figure}[b]
    \begin{center}
     \includegraphics[width=8.4cm]{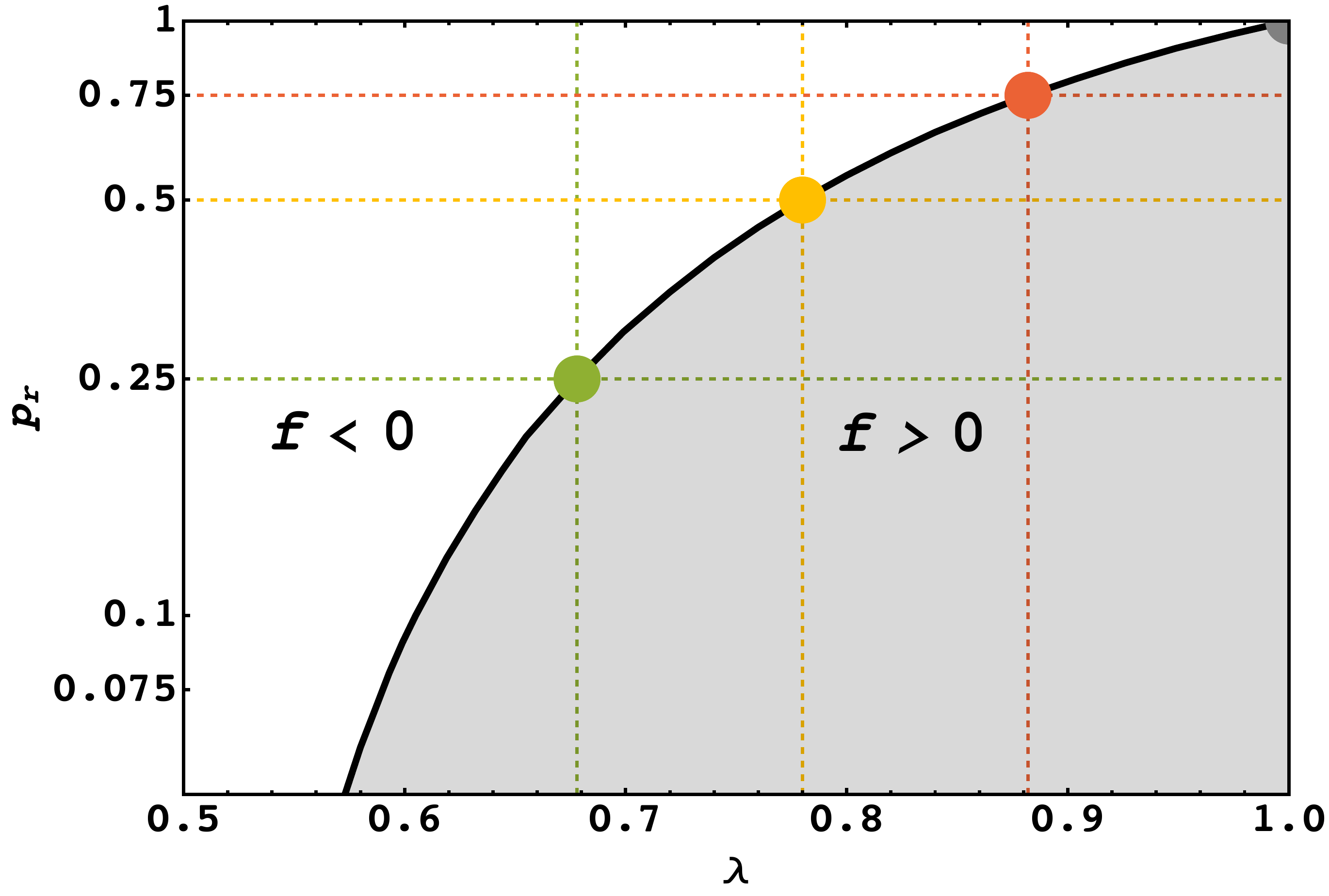}
    \end{center}
    \caption{A phase diagram of $p_r$ vs. $\lambda$ based on the qualitative effect of resetting on gated drift-diffusion. The black line represents the condition for resetting transition ($\lambda_c$), which divides the entire phase space in two parts. For $\lambda<\lambda_c$ resetting expedites transport (white regime), whereas for $\lambda\geq\lambda_c$ resetting fails to expedite transport (gray regime). Here we take $D=1$, $x_0=2$ and $\alpha=0.5$ [i.e., $p_r=1/(2\beta+1)$]  for all cases. For the ungated process ($p_r=1$), $\lambda_c=1$ and it decreases with $p_r$. The colored discs on the black line present the cases shown in \fref{Fig4}. Following the analysis of $\langle T_{r}^G\rangle$ in the limit $r\to 0$ [see \eref{expf} and \eref{separatrix}], the separatrix (black line) is also obtained by plotting $f=0$. Resetting is beneficial for $f<0$ (white regime), but not for $f>0$ (gray regime). }
      \label{Fig5}
\end{figure}
\indent
In order to better understand how the critical value of $\lambda$ changes with the reactive occupancy of the target, next we numerically calculate $\lambda_c$ as a function of $p_r$. In \fref{Fig5}, we construct a phase diagram spanned by the reactive occupancy $p_r$ and the drift velocity $\lambda$, where $\lambda_c$ acts as the separatrix that divides the entire phase space in two parts; one where resetting expedites first-passage (white regime) and the other where it can not (gray regime). We can recover the core results observed in \fref{Fig4} from \fref{Fig5} in the following way. For each value of $p_r\in [0,1]$ [encountered by moving horizontally through \fref{Fig5}], the dynamics is diffusion-controlled and resetting is beneficial if $\lambda<\lambda_c$, whereas the dynamics is drift-controlled and resetting turns out to be non-beneficial if $\lambda\geq\lambda_c$. It is clear from \fref{Fig5} as well that with increase in reactive occupancy, $\lambda_c$ increases to finally become unity for the ungated process ($p_r=1$). \fref{Fig5} is, therefore, a complete yet compact representation of the condition for resetting to expedite gated drift-diffusion. Next, we discuss an alternative approach that can successfully generate this phase diagram by carefully exploring the resetting criterion in the limit $r\to 0$.\\
\indent
Revisiting \fref{Fig3}, we see that it clearly indicates when initial (in the limit $r\to 0$) slope of the $\langle T_{r}^G \rangle $ vs. $r$ curve is negative, introduction of resetting proves itself beneficial by decreasing the average MFPT. In contrast, a positive slope of such curve in the limit $r\to 0$ suggests that introduction of resetting increases the average MFPT there. This motivates us to explore the condition for resetting to accelerate gated drift-diffusion by analyzing the expression of the average MFPT in the limit $r\to 0$. To do that, we first expand $\langle T_{r}^G \rangle$ in $r$ for an infinitesimal resetting rate to obtain
\bal
\langle T_{r}^G \rangle 
\approx  \langle T^G \rangle 
   +r \left[\frac{\partial\langle T_r^G \rangle}{\partial r} \right]_{r=0}\mbox{\hspace{-0.5cm}}+ O(r^2),
  \label{expf}
\eal
where $\langle T^G \rangle $ is the average MFPT in the absence of resetting. Following \eref{mfpt_av}, we get [see \aref{appc} for an alternative derivation]
\bal
\langle T^G \rangle = \frac{1}{\lambda}\left[x_0+\frac{2 \beta D}{\alpha \left(\sqrt{4 \alpha 
   \text{D}+4 \beta  \text{D}+\lambda ^2}-\lambda \right)}\right].
  \label{tg}
\eal
Note that in the absence of gating ($p_r=1$, or $\beta=0$), \eref{tg} reduces to $\langle T\rangle = x_0/\lambda$ \cite{REDNER}, which implies that $\langle T^G\rangle >\langle T\rangle$, since the second term in the right hand side of \eref{tg} is always positive.\\
\indent
The expression for the second term in the right hand side of \eref{expf} is fairly complicated, here we write it in a simpler form by introducing some new parameters, viz., $\gamma\coloneqq \lambda^2/2D$, $Pe\coloneqq x_0 \lambda/2 D$, and $\kappa\coloneqq\sqrt{1+[4D(\alpha+\beta)/\lambda^2]}-1$, such that
\begin{align}
   f\coloneqq \left.\frac{\partial\langle T_r^G \rangle}{\partial r} \right|_{r=0}\mbox{\hspace{-0.4cm}}=\frac{\beta}{2\gamma^2 \alpha }\Big[\frac{2 Pe -1}{\kappa}-&\frac{2}{\kappa^2(\kappa+1)}+\frac{2 \gamma e^{-\kappa Pe}}{\kappa (\alpha+\beta)}\Big] \nonumber \\
     +&\frac{1}{2\gamma^2}Pe(Pe-1).
     \label{separatrix}
 \end{align}
Note that $Pe$ in \eref{separatrix} is the P\'eclet number, i.e., the ratio between the rate of driven transport to that of diffusive transport and $\gamma^{-1}$ is the fastest first-passage time (smallest decay time) in the strong drift limit as was pointed out by Redner in \cite{REDNER}. \\
\indent
It is evident from \eref{expf} that in the limit $r \to 0$, resetting is expected to expedite the completion of the process, i.e., $\langle T_{r}^G \rangle < \langle T^G \rangle$, when $f<0$. This is a sufficient condition (may not be necessary) for resetting to be useful. When $f>0$, however, resetting is expected to delay the completion of the process, i.e., $\langle T_{r}^G \rangle > \langle T^G \rangle$. Therefore, the condition $f=0$ should divide the entire phase space created by ($\lambda,p_r$) in two parts; one where resetting is beneficial ($f<0$) and the other where it is not ($f>0$). Indeed, when we plot $f=0$ following \eref{separatrix} in the same phase diagram presented in \fref{Fig5}, it exactly overlaps on the existing separatrix plotted earlier by calculating the critical drift $\lambda_c$ as a function of $p_r$. Therefore, the condition $f<0$ marks the phase where introduction of resetting accelerates transport to the target (white regime) while $f>0$ marks the phase where introduction of resetting delays the same (gray regime), as expected. Revisiting \eref{separatrix}, we note that in the limit of $p_r \to 1$ (for $\beta\to 0$), the first term of \eref{separatrix} vanishes. Then, $f<0$ boils down to $Pe<1$ (meaning $Pe=1$ is the resetting transition point), the condition where resetting accelerates ungated drift-diffusion, as obtained in earlier works \cite{branching,PECLET,local}.  \\
\label{various-alpha}
The phase diagram in \fref{Fig5} is generated only for a fixed value for $\alpha$. Next, we perform similar exercise for other values of $\alpha$ and display the results in \fref{Fig6}. It is evident from \fref{Fig6} that the separatrix, i.e, the phase boundary that separates the two phases -- the ``resetting-beneficial'' phase at the left and the ``resetting-detrimental'' at the right -- varies with $\alpha$. In fact, for larger values of $\alpha$, the separatrix divides the phases in such a way that the  ``resetting-beneficial'' phase becomes considerably smaller and the ``resetting-detrimental'' phase occupies most of the phase space, implying that the effect of resetting becomes rather constrained in that limit. Next, we examine two important limits of the gating rates, viz., $\alpha \to 0$ and $\alpha, \beta \to \infty$. 
\begin{figure}[t]
    \centering
    \includegraphics[width=8.0cm]{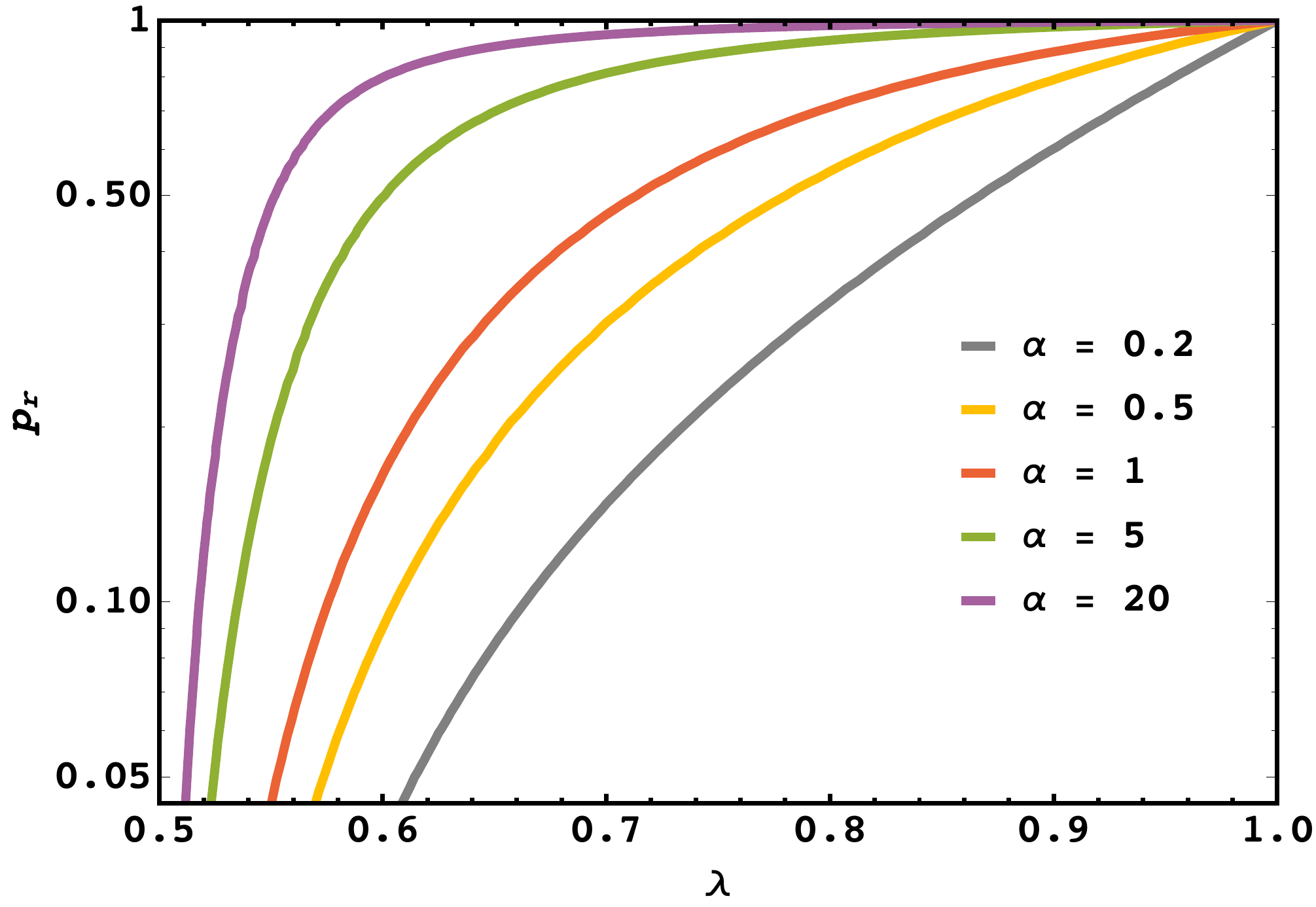}
    \caption{A phase diagram of $p_r$ vs. $\lambda$ for different values of $\alpha$. Each phase boundary (obtained for a certain value of $\alpha$, shown by the curves) divides the phase space into a ``resetting-beneficial'' phase (left to the curve) and a ``resetting-detrimental''  phase (right to the curve). For large values of $\alpha$, the ``resetting-beneficial'' phase becomes considerably smaller as the ``resetting-detrimental'' phase occupies the majority of the phase space. }
    \label{Fig6}
\end{figure}
\subsection{The limit $\alpha \to 0$}
\label{limits-of-alpha}
\begin{figure}[t]
    \centering
    \includegraphics[width=8.25cm]{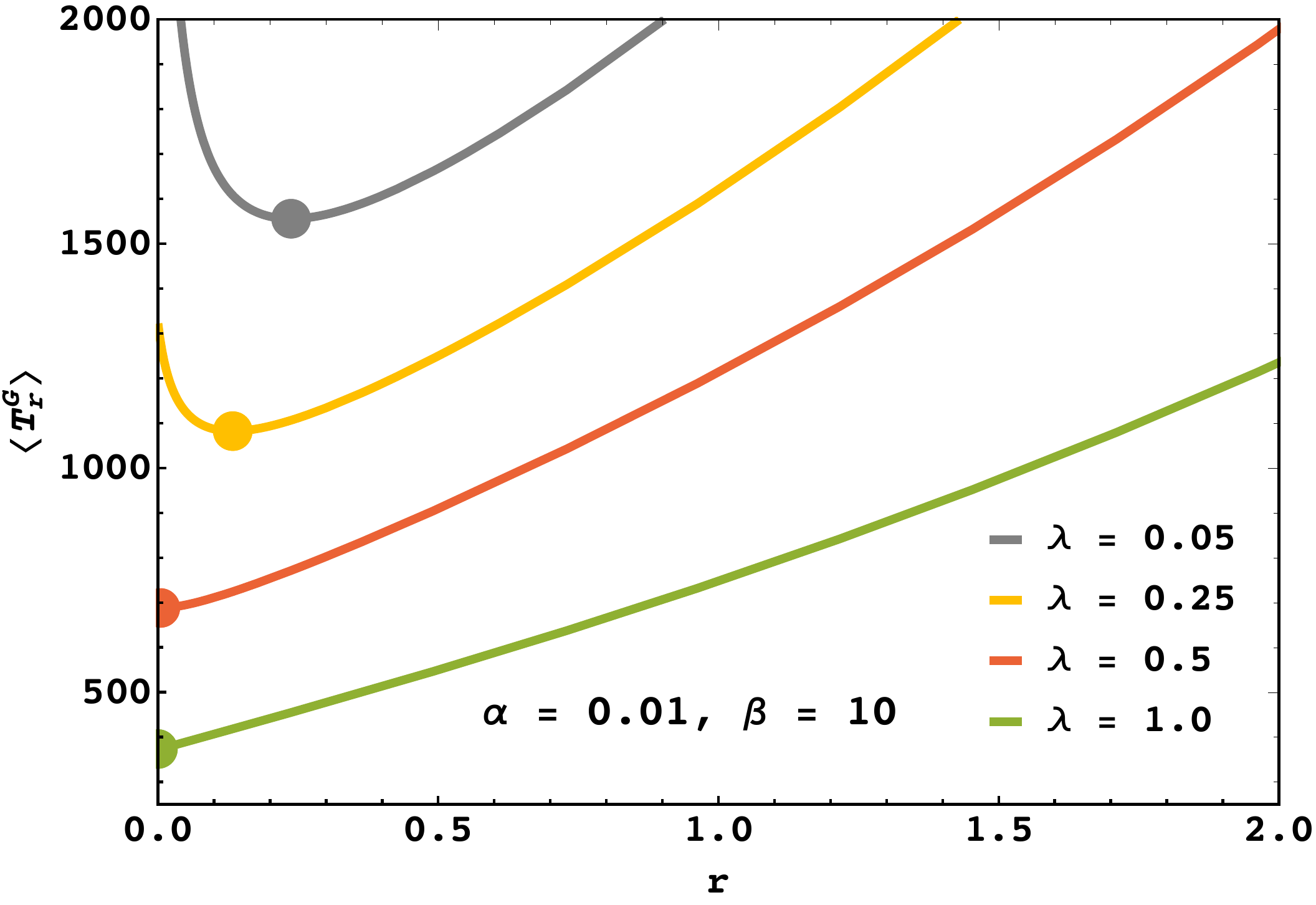}
    \caption{Monotonic and non-monotonic behaviour of the MFPT in the limit $\alpha \to 0$ (i.e., $\beta/\alpha \gg 1$), as obtained from \eref{mfpt_al}. The non-monotonic behaviour gradually vanishes as $\lambda$ goes beyond the resetting transition point $\lambda_c=D/x_0=0.5$. This can be obtained by setting $r^{\star}$ in \eref{orr_lim} to zero.}
    \label{Fig7}
\end{figure}

The limit $\alpha \to 0$ essentially implies that the target is reflective almost all the time. In what follows, we will show that this is a delicate limit and should be handled carefully. Strictly speaking, the limit $\alpha \to 0$ should be interpreted as $\beta \gg \alpha$, which means that the target has higher probability to remain non-reactive than being reactive. Using this limit in \eref{mfpt_av}, one finds
    \bal
\langle T_{r}^G\rangle=\frac{1}{r}\left(e^{\mu_1 x_0}-1\right)+
\frac{\beta \mu_1}{r \alpha \mu_2} e^{\mu_1 x_0},
\label{mfpt_limit}
\eal
where $\mu_{1}=(-\lambda +\sqrt{\lambda ^2+4 D r})/2D>0$ and $\mu_{2}=(-\lambda +\sqrt{\lambda ^2+4 D (\alpha +\beta+r )})/2D>0$, as before. Moreover, one can disregard the first term in the right hand side of \eref{mfpt_limit} for finite resetting rate $r$ (assuming $\beta \gg r$) to find 
\begin{align}
  \langle T_{r}^G\rangle= \frac{\beta  \left(\sqrt{4 D r+\lambda ^2}-\lambda \right) e^{\frac{x_0 \left(\sqrt{4 D r+\lambda ^2}-\lambda \right)}{2 D}}}{\alpha  r \left(\sqrt{4 \beta  D+\lambda ^2}-\lambda \right)}.
  \label{mfpt_al}
\end{align}
To understand the behavior of the average MFPT, we plot $\langle T_{r}^G\rangle$ as a function of the resetting rate in \fref{Fig7} for large $\beta$ and small $\alpha$ keeping $\beta/\alpha \gg 1$. Intriguingly, we find that in this limit, $\langle T_{r}^G\rangle$ shows both monotonic and non-monotonic behavior as one varies $r$ for different $\lambda$. In particular, the latter case implies that there exists a resetting rate for which $\langle T_{r}^G\rangle$ becomes optimally minimum.
To find this optimal resetting rate, we set $\left.\frac{d \langle T_{r}^G\rangle }{dr}\right|_{r=r^{\star}}=0$ and obtain 
\begin{align}
    r^{\star}=\frac{D}{x_0^2}-\frac{\lambda}{x_0}, \label{orr_lim}
\end{align}
which is a simple function of the diffusion constant, initial position and drift. The linear behavior of $r^{\star}$ with respect to the drift variable $\lambda$ is also noteworthy. Finally, setting $\lambda=0$, one recovers $ r^{\star}=D/x_0^2$ which was obtained in \cite{gated-1}.\\
\indent
The appearance of an ORR in the limit $\alpha\to0$ i.e., when the target is poorly reactive (the so-called \textit{cryptic} regime \cite{gated-0,gated-2}) is counter-intuitive in contrast to the case of $\alpha=0$ (purely reflective boundary). It can be argued that a diffusing particle usually makes several encounters (also aided by the drift) with the target regardless of the target's state. In the current context, although most of the time the particle may remain unsuccessful to find the target in a reactive state, it can still get absorbed whenever there is an element of chance for the target to become reactive. More details about such cryptic targets and their natural appearances in chemical, biological and ecological systems can be found in \cite{gated-0,gating-review,golding,prey-cryptic}.\\
\subsection{The limit $\alpha,\beta\to\infty$}
\label{limits-of-alpha-beta}
In the limit $\alpha,\beta\to\infty$, recalling \eref{separatrix} we can approximate $\kappa\approx \sqrt{4D(\alpha+\beta)}/\lambda$. Thus, one can safely neglect the terms of order $O(\frac{1}{\kappa^2})$ including the exponential term. Under this assumption, setting $f=0$ in  \eref{separatrix} we obtain an exact expression for the critical drift $\lambda_c$ that reads
\begin{align}
     \lambda_c=\frac{2 D}{x_0}\left[\frac{   D+ x_0 \frac{\alpha}{\beta}\sqrt{\alpha+\beta}\sqrt{D}}{2D+ x_0 \frac{\alpha}{\beta}\sqrt{\alpha+\beta}\sqrt{D}}\right].
      \end{align}
We notice that even in this limit $p_r$ does not solely control the dynamics of the system. More from the physical point of view, one can refer to this limit as the partially reactive boundary, where the target switches between the reactive and non-reactive states so fast (i.e., the time scale of such switch is much smaller compared to the time scale of drift-diffusion) that the particle feels an average state of the target and interacts with it with a certain probability all the time -- see also \cite{gating-review,gated-0}. \\
\indent
So far, we performed a detailed analysis to understand the effect of resetting on the dynamics of gated drift-diffusion. In the next section, we turn our attention to the maximal speed-up that can be gained by resetting the process at an optimal rate. 
\section{The maximal speedup for process completion}
\label{speed-up}
\begin{figure}[b]
    \begin{center}
      \includegraphics[width=8.2cm]{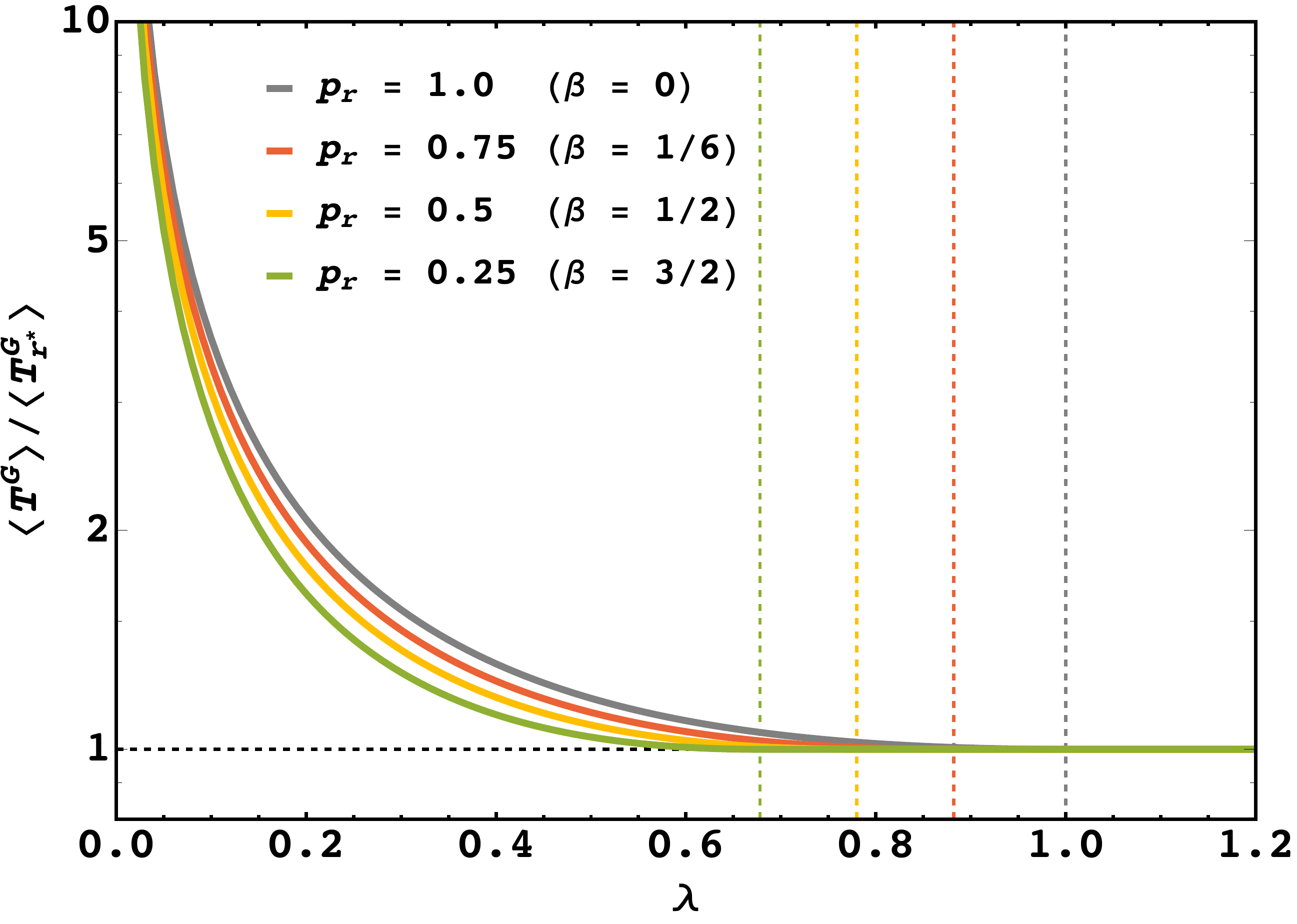}
    \end{center}
    \caption{The maximal speedup for the gated process with resetting compared to the gated process without resetting, $\left<T^G\right>/\left<T^G_{r^{\star}}\right>$, vs the drift velocity $\lambda$, for different values of the reactive occupancy, $p_r$. The vertical dashed lines mark the points of resetting transitions for curves of same color, denoted $\lambda_c$, such that $\left<T^G\right>/\left<T^G_{r^{\star}}\right>>1$ for $\lambda<\lambda_c$, and $\left<T^G\right>/\left<T^G_{r^{\star}}\right>=1$ at $\lambda\geq\lambda_c$. We take $D=1$, $x_0=2$ and $\alpha=0.5$ [i.e., $p_r=1/(2\beta+1)$] for all cases. In the absence of gating ($p_r=1$, gray curve), the resetting transition is observed at $\lambda_c=2D/x_0=1$, and $\lambda_c$ decreases when gating is introduced ($p_r<1$).}
      \label{Fig8} 
      \end{figure}

The maximal speedup, i.e., the speedup rendered by optimal resetting rate $r^{\star}$, is generally defined as the ratio of the MFPT of the original (underlying) process without resetting to that of the process with optimal resetting. Therefore, setting $r=r^{\star}$ in \eref{mfpt_av} and utilizing \eref{tg}, we obtain the maximal speedup for {\it gated} drift-diffusion process in a straightforward way, which reads
\begin{eqnarray}
\footnotesize
\frac{\langle T^G \rangle}{\langle T^G_{r^{\star}} \rangle}=
\begin{cases}
\frac{r^{\star} (\alpha+\beta)(2D\beta+x_0\alpha\mu_2^{\star})}{\lambda\left[r \beta\mu_1^{\star} e^{\frac{x_0}{2D}\left(\mu_1^{\star}+\mu_2^{\star}\right)}+\left(\alpha+\beta\right)\left[e^{\frac{x_0\mu_1^{\star}}{2D}}{(\beta \mu_1^{\star}+\alpha\mu_2^{\star})-\alpha\mu_2^{\star}}\right]\right]}\;\;\mbox{for}\;\; \lambda<\lambda_c  \\
1\;\;\;\;\;\;\;\;\;\;\;\mbox{for}\;\; \lambda\geq\lambda_c,
\end{cases}
\label{speedup_lambda_c}
\end{eqnarray}
where $\mu_{1}^{\star}=\mu_1(r^{\star})$ and $\mu_{2}^{\star}=\mu_2(r^{\star})$, $\mu_1$ and $\mu_2$ having the expressions given after \eref{mfpt_av}. In \fref{Fig8}, we plot $\langle T^G \rangle/\langle T^G_{r^{\star}} \rangle$ for different values of the reactive occupancy $p_r$. \fref{Fig8} indicates that the maximal speedup for gated drift-diffusion is most marked when drift towards the target is negligible. With increase in $\lambda$, it gradually decreases to finally become unity at the point of resetting transition, $\lambda_c$. We see that when $p_r=1$, i.e, in the absence of gating, the resetting transition is observed at $\lambda_c=1$ for our choice of parameters \cite{PECLET}, but when $p_r$ is decreased below unity, the transition is observed for lower values of $\lambda_c$, as observed earlier in \fref{Fig5}.\\
\indent
Next, we compare $\left<T^G_{r^{\star}}\right>$ to $\left<T\right>$, the MFPT for {\it ungated} drift-diffusion without resetting, to explore whether it is possible to overcome the increase in the MFPT due to gating by resetting the process in an optimal way. Recalling that the MFPT of ungated drift-diffusion is given by $\langle T\rangle = x_0/\lambda$ and utilizing \eref{mfpt_av} for $r=r^{\star}$ as before, we get 
\begin{eqnarray}
\footnotesize
\frac{\langle T\rangle}{\langle T^G_{r^{\star}} \rangle}=
\begin{cases}
\frac{r^{\star}x_0\alpha(\alpha+\beta)\mu_2^{\star}}{\lambda\left[r \beta\mu_1^{\star} e^{\frac{x_0}{2D}\left(\mu_1^{\star}+\mu_2^{\star}\right)}+\left(\alpha+\beta\right)\left[e^{\frac{x_0\mu_1^{\star}}{2D}}{(\beta \mu_1^{\star}+\alpha\mu_2^{\star})-\alpha\mu_2^{\star}}\right]\right]}\;\;\mbox{for}\;\;\;\;\lambda<\lambda_c  \\
\frac{x_0\alpha\mu_2^{\star}}{2D\beta+x_0\alpha\mu_2^{\star}}\;\;\;\;\;\;\;\;\;\;\;\mbox{for}\;\; \lambda\geq\lambda_c.
\end{cases}
\label{speedup_lambda_ug}
\end{eqnarray}
\begin{figure}[t]
    \begin{center}
      \includegraphics[width=8.2cm]{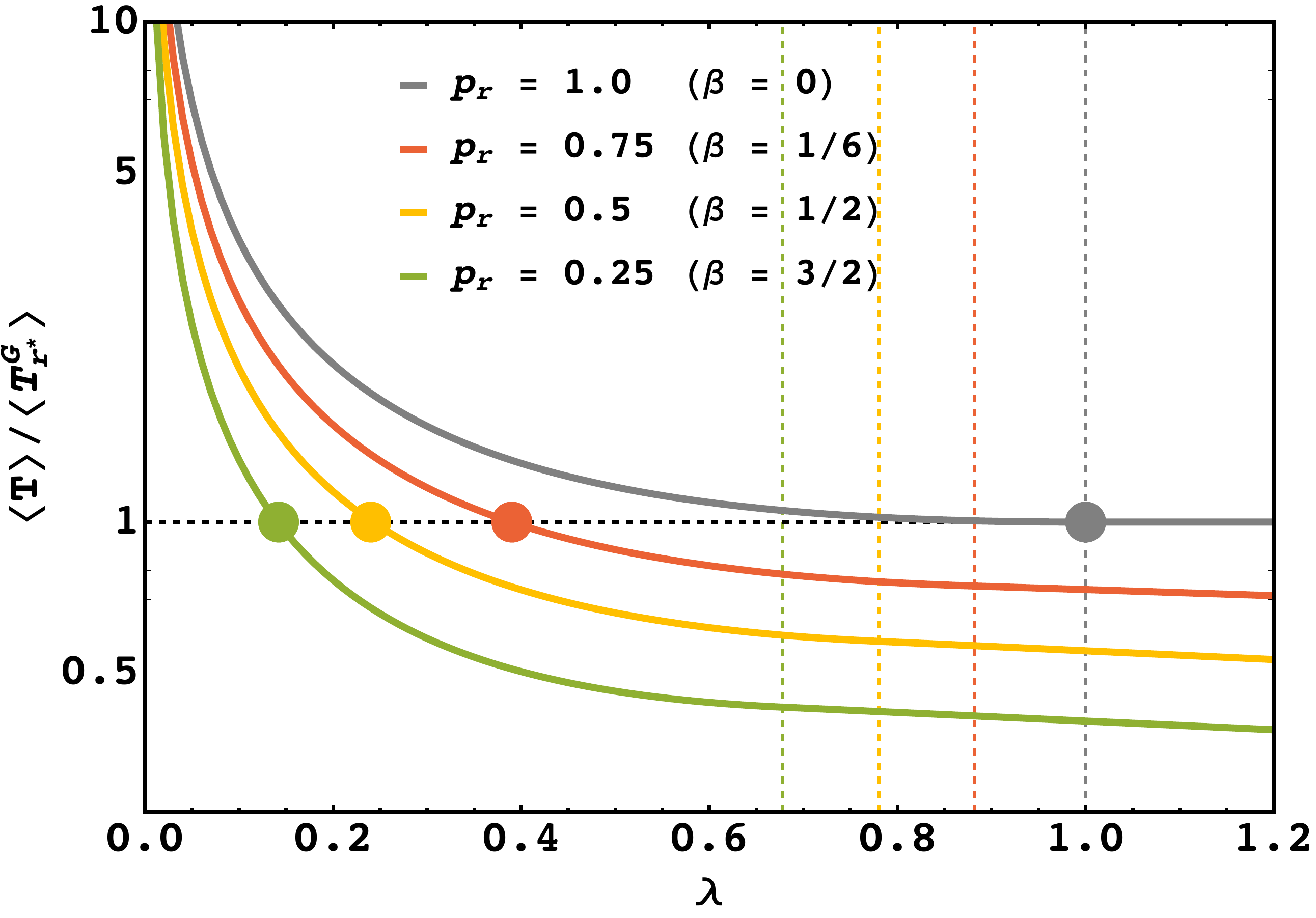}
    \end{center}
    \caption{The maximal speedup for the gated  process with resetting compared to the ungated process without resetting, $\left<T\right>/\left<T^G_{r^{\star}}\right>$, vs $\lambda$, for different values of $p_r$. The vertical dashed lines mark the points of resetting transition ($\lambda_c$) for curves of same color, whereas the colored discs mark the values of $\lambda_c^0$, such that optimal resetting expedites gated process beyond the original ungated process only when $\lambda<\lambda_c^0$. Here $D=1$, $x_0=2$ and $\alpha=0.5$ [i.e., $p_r=1/(2\beta+1)$] for all cases. In the absence of getting ($p_r=1$, gray curve), the point of resetting transition ($\lambda_c=2D/x_0=1$) coincides with $\lambda_c^0$.}
      \label{Fig9}
\end{figure}
Plotting $\langle T\rangle/\langle T^G_{r^{\star}}\rangle$ as a function of $\lambda$ in \fref{Fig9} for different values of $p_r$, we see that the maximal speedup compared to the original ungated process is infinite when there is no drift towards the target and it decreases with an increase in $\lambda$, as expected. In fact, \fref{Fig9} clearly shows that for sufficiently low values of $\lambda$, optimal resetting can make the process even $>10$ times faster! It proves that resetting is indeed a useful strategy to compensate for the delay due to gating, and when the dynamics is diffusion-controlled, it can even improve the rate of transport (which is inversely proportional to the mean completion time) beyond the original ungated process without resetting. It is also observed from \fref{Fig9} that in the absence of gating (when $p_r=1$), the minimal possible value for $\langle T\rangle/\langle T^G_{r^{\star}}\rangle$ is unity, which is achieved for $\lambda\geq \lambda_c$\cite{PECLET}. In contrast, when $p_r<1$, $\langle T\rangle/\langle T^G_{r^{\star}}\rangle$ is reduced below unity even before the resetting transition sets in. Therefore, denoting $\lambda_c^0$ as the critical value of $\lambda$ [marked in \fref{Fig9} by the colored discs] where $\langle T\rangle/\langle T^G_{r^{\star}}\rangle$ becomes unity for a certain $p_r$, we observe that $\lambda_c^0\leq \lambda_c$. The equality holds only for the ungated process and the difference between $\lambda_c^0$ and $\lambda_c$ is prominent for lower values of $p_r$. 
These observations lead us to identify all the possible distinct regimes where resetting can benefit the completion of the gated drift-diffusion process. In what follows, we construct a comprehensive phase diagram in the parameter space encapsulating all these effects of resetting.
\section{The complete phase diagram}
\label{full-phase}
\begin{figure}[t]
    \begin{center}
      \includegraphics[width=8.5cm]{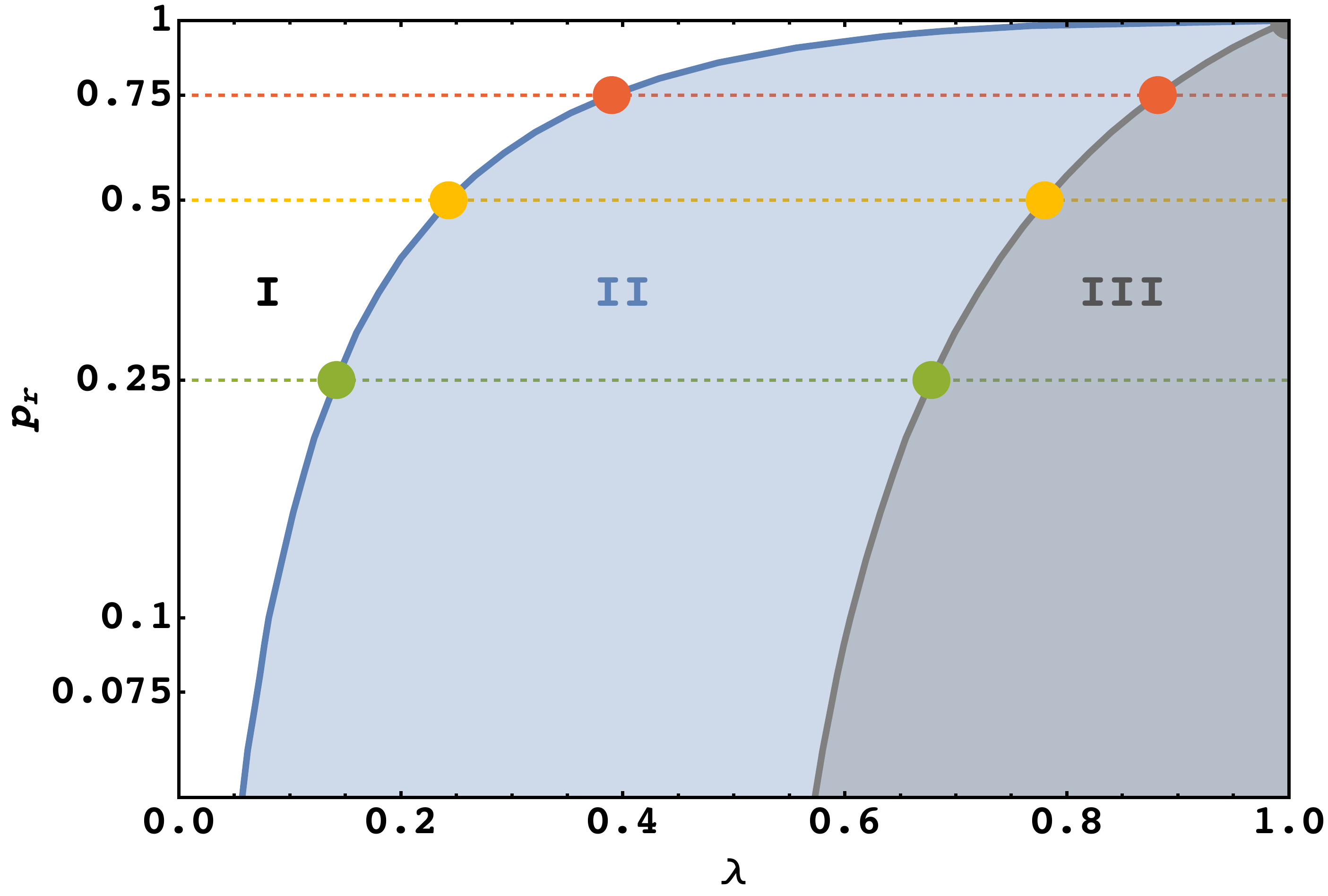}
    \end{center}
    \caption{A complete phase diagram of $p_r$ vs. $\lambda$ showing three distinct phases; (i) phase I: where optimal resetting enhances the rate of gated drift-diffusion beyond the original (without resetting) ungated process, given by $\left<T^G_{r^{\star}}\right><\left<T\right><\left<T^G\right>$, (ii) phase II: when optimal resetting improves the rate of the gated process, but not compared to the original ungated process, given by $\left<T\right>\leq\left<T^G_{r^{\star}}\right><\left<T^G\right>$, and (iii) phase III: when resetting can not improve the rate of   gated process, given by $\left<T\right><\left<T^G\right>\leq\left<T^G_{r^{\star}}\right>$. The horizontal colored lines mark the cases shown in \fref{Fig9}. Here we consider $\alpha=0.5$ [such that $p_r=1/(2\beta+1)$], $D=1$ and $x_0=2$. Phases I and II merge together at $\lambda_c=2D/x_0=1$, the point of resetting transition for drift-diffusion in the absence of gating ($p_r=1$). }
      \label{Fig10}
\end{figure}
To construct the complete phase diagram for the problem, we revisit \fref{Fig5} and \fref{Fig8}, and recall that  the entire phase space [spanned by $(\lambda,p_r)$] is divided into two parts, viz., $\langle T^G_{r^{\star}}\rangle<\langle T^G\rangle$ [i.e., where optimal resetting expedites transport] and $\langle T^G_{r^{\star}}\rangle\geq\langle T^G\rangle$ [i.e., where optimal resetting can not expedite transport], and the transition between these two phases takes place at a critical point $\lambda=\lambda_c$. The study of the maximal speedup of the gated process with resetting compared to the ungated process without resetting in Section IV suggests that we can further divide the former phase in two parts: one where the rate of transport for the gated process with optimal resetting is higher than that of original ungated process i.e., $\langle T^G_{r^{\star}}\rangle<\langle T\rangle$ and the other where it is not i.e., $\langle T^G_{r^{\star}}\rangle\geq \langle T\rangle$. The transition between these two phases happens at $\lambda=\lambda_c^{0}\leq\lambda_c$. Since gating essentially makes drift-diffusion slower in the absence of resetting, i.e, $\left<T\right><\left<T^G\right>$ [see \eref{tg}], summarizing our observations, we construct a complete phase diagram for the present problem [displayed in \fref{Fig10}], which consists of three distinct phases: (i) phase I: when optimal resetting makes the gated process faster than the original ungated (and hence the original gated) process, given by $\left<T^G_{r^{\star}}\right><\left<T\right><\left<T^G\right>$, (ii) phase II: when optimal resetting makes the gated process faster than the original gated process, but not the original ungated process, given by $\left<T\right>\leq\left<T^G_{r^{\star}}\right><\left<T^G\right>$, and (iii) phase III: when resetting can not make the gated process faster than the original gated (and hence the original ungated) process, given by $\left<T\right><\left<T^G\right>\leq\left<T^G_{r^{\star}}\right>$. The transition points $\lambda_c^0$ create the separatrix between phases I and II, whereas the transition points $\lambda_c$ create that between phases II and III. Since $\lambda_c^0=\lambda_c$ for $p_r=1$, in \fref{Fig10} we find phases I and II to merge together in the absence of gating. This diagram in the phase space of two important parameters of the system, namely the reactive occupancy and the bias, allows us to delineate the exact nature of resetting in the process completion. In other words, we can gain maximal benefits 
from a precise and \textit{a priori} knowledge of the parameter space.
\section{Conclusions}
\label{summary}
\noindent
In this work, we performed an in-depth analysis on the completion time statistics of drift-diffusive transport to a stochastically gated target in the presence of Poissonian resetting. In particular, we strategically explored the conditions where resetting can enhance the rate of such transport as has been shown for ungated processes where resetting stabilizes the non-equilibrium motion \cite{Restart1,Evansrev2020,Restart4,ss,work-fluc} by removing the detrimental long trajectories that result in a slower transport rate. Projecting the general problem of gated drift-diffusion with resetting to a gated chemical reaction initiated by a catalyst [as discussed in \fref{Fig1}], the main results of the present work can be interpreted as follows. \\
\indent 
We observed that the rate of product formation depends on an interesting interplay between the chemical potential drive that governs the reaction ($\lambda$), the probability ($p_r$) that gated reactant stays in its activated state [when $R_2$ exists as $R_2^{\star}$], and the rate of unbinding (or resetting, with rate $r$) of the catalyst $C$ from $CR_1$. When the chemical potential drive towards the product ($\lambda>0$) is low/moderate such that the reaction is diffusion-controlled, the rate of reaction is maximized for an optimal unbinding rate ($r^{\star}$) of the catalyst. In contrast, when the drive $\lambda$ is strong, i.e., the reaction is drift-controlled, unbinding of the catalyst $C$ from $CR_1$ decreases the rate of reaction. A transition is thus observed at a critical drive, $\lambda_c$, which grows with $p_r$ and becomes maximum for $p_r=1$, i.e., for the ungated reaction. Strikingly enough, we observed that for $\lambda<\lambda_c^0$ [when $\lambda_c^0\leq\lambda_c$ is another critical value of $\lambda$ that increases with $p_r$ and attains a maximum $\lambda_c^0=\lambda_c$ at $p_r=1$], optimal unbinding with a rate $r^{\star}$ can make the reaction even $>10$ times faster compared to the ungated reaction in the limit $r\to 0$ (i.e., where the binding step is almost irreversible).\\
\indent
These observations lead to a complete phase diagram based on the {\it qualitative} and {\it quantitative} effect of optimal unbinding (resetting) on gated chemical reaction [modeled by drift-diffusion to a gated target] that consists of three distinct phases. Recalling that \cite{effect-pot-3}
rate of product formation $\propto \mbox{(mean completion time of reaction)}^{-1}$, these three phases are identified through the following conditions: (i) where $\langle T^G_{r^{\star}}\rangle < \langle T\rangle < \langle T^G \rangle$, i.e., where the rate of gated chemical reaction is enhanced by optimal unbinding of catalyst beyond that of ungated/gated reactions when the binding is almost irreversible [$r\to 0$], (ii) where $\langle T\rangle \leq \langle T^G_{r^{\star}}\rangle < \langle T^G\rangle$, i.e., when optimal unbinding improves the rate of gated reaction, but not beyond the ungated reaction in the limit $r\to 0$, and (iii) where $\langle T\rangle <\langle T^G\rangle \leq \langle T^G_{r^{\star}}\rangle$, i.e., when unbinding fails to make the gated reaction faster than either of the gated/ungated reaction for almost irreversible binding.\\
\indent
The model considered herein generally applies to gated drift-diffusion under the influence of resetting. The major advantage besides its  analytical tractability is that one can also gain deep insights about the intricate trade-offs between gating and resetting mechanisms, both of which are essential components of chemical reaction networks. Generalization of this simple model to a generic space-dependent diffusion process in an arbitrary energy landscape in the presence of gated targets would be a potential research avenue. A detailed numerical analysis to this end would be a worthwhile pursuit. Notably, such theoretical models can capture physical situations arising in experiments that study, e.g., completion time statistics of protein folding by gated fluorescence quenching. There, the protein is tagged/labelled by a fluorophore reversibly binds the protein (here unbinding is similar to resetting) to impart fluorescence properties \cite{Quenching}. Once the tagged protein folds to its native state, the quencher selectively reacts with the active site of that folded protein in its fluorescent state, provided that site is in its open (exposed) conformation. If the active site of the folded protein remains in a closed (hidden) conformation, the quencher fails to react with it, which implicates gating. A successful reaction thus occurs only in the exposed conformation, which leads to subsequent quenching of fluorescence thereby marking the completion of the folding process \cite{Gated_Quenching}. We believe that our work can shed light in understanding and harnessing various gating and resetting protocols inherent to such systems.
\vspace{-0.4cm}
\section*{Acknowledgement}
\noindent  AP gratefully acknowledges research support from the DST-SERB Start-up Research Grant Number SRG/2022/000080 and the Department of Atomic Energy, Govt. of India. DM thanks SERB (Project No. ECR/2018/002830/CS), DST, Govt. of India, for financial support and IIT Tirupati for the new faculty seed grant. SR acknowledges the Elizabeth Gardner Fellowship by School of Physics \& Astronomy, University of Edinburgh and the INSPIRE Faculty research grant by DST, Govt. of India, executed at IIT Tirupati. The numerical calculations reported in this work were carried out on the Nandadevi cluster, which is maintained and supported by IMSc’s High-Performance Computing Center. For the purpose of open access, the authors have applied a Creative Commons Attribution (CC BY) licence to any Author Accepted Manuscript version arising from this submission. We thank the anonymous Reviewers for their insightful remarks. 
\vspace{-0.25cm}
\section*{DATA AVAILABLITY}
\noindent
The data that supports the findings of this work are available within the paper, its appendices, and in \cite{github}.\\

\appendix
\vspace{-0.35cm}
\section{Calculation of the average MFPT by solving \eref{bfpe} in the Laplace space}
\renewcommand{\theequation}{A.\arabic{equation}}
\setcounter{equation}{0}
\label{appa}
Here, we provide the solution of \eref{bfpe} in the Laplace space. Following that, we calculate the mean first passage time for a diffusing particle with drift,  starting from an initial position $x_0$, to reach the gated target. We start by Laplace transforming \eref{bfpe}
\begin{align}
D\frac{\partial^2 \tilde{Q}_{0}(s|x_0)}
{\partial x_0^2}-\lambda \frac{\partial \tilde{Q}_{0}(s|x_0)}{\partial x_0}
&-(s+\alpha+r)\tilde{Q}_{0}(s|x_0)+\alpha
\tilde{Q}_{1}(s|x_0)\nonumber\\
&=-1-r\tilde{Q}_{0}(s|x_r)
\nonumber\\
D\frac{\partial^2 \tilde{Q}_{1}(s|x_0)}
{\partial x_0^2}-\lambda \frac{\partial \tilde{Q}_{1}(s|x_0)}{\partial x_0}
&-(s+\beta+r)\tilde{Q}_{1}(s|x_0)+\beta
\tilde{Q}_{0}(s|x_0)\nonumber\\
&=-1-r\tilde{Q}_{1}(s|x_r),
    \label{fpe_lt}
\end{align}
where $\tilde{Q}_{\sigma}(s|y)\coloneqq\int_0^{\infty}dt\;e^{-st}Q_{\sigma}(t|y)$ denote the Laplace 
transform of $Q_{\sigma}(t|y)$. \eref{fpe_lt} is a second order, linear and non-homogeneous differential equation. Considering 
\begin{equation}
\tilde{Q}_{\sigma}(s|x_0)=\tilde{Q}_{\sigma}^h(s|x_0)+\tilde{Q}_{\sigma}^{inh}(s|x_0),
\label{Q_h_inh}
\end{equation}
where $\tilde{Q}_{\sigma}^h(s|x_0)$ and $\tilde{Q}_{\sigma}^{inh}(s|x_0)$ denote the homogeneous and inhomogeneous parts of $\tilde{Q}_{\sigma}(s|x_0)$, respectively,
we can rewrite \eref{fpe_lt} in two separate parts. The homogeneous part reads
\begin{align}
D\frac{\partial^2 \tilde{Q}_{0}^h(s|x_0)}
{\partial x_0^2}-\lambda \frac{\partial \tilde{Q}_{0}^h(s|x_0)}{\partial x_0}-(s+\alpha&+r)\tilde{Q}_{0}^h(s|x_0)  \nonumber \\
&+\alpha \tilde{Q}_{1}^h(s|x_0)=0, \nonumber\\
D\frac{\partial^2 \tilde{Q}_{1}^h(s|x_0)}
{\partial x_0^2}-\lambda \frac{\partial \tilde{Q}_{1}^h(s|x_0)}{\partial x_0} -(s+\beta&+r)\tilde{Q}_{1}^h(s|x_0)  \nonumber \\
&+\beta
\tilde{Q}_{0}^h(s|x_0)=0.
    \label{fpe_lth}
\end{align}
Similarly, the inhomogeneous part reads
\begin{align}
-(s+\alpha+r)\tilde{Q}_{0}^{inh}(s|x_0)+\alpha
\tilde{Q}_{1}^{inh}(s|x_0)=-1-r\tilde{Q}_{0}(s|x_r),\nonumber\\
-(s+\beta+r)\tilde{Q}_{1}^{inh}(s|x_0)+\beta
\tilde{Q}_{0}^{inh}(s|x_0)=-1-r\tilde{Q}_{1}(s|x_r), \nonumber \\
\label{Q_inh}
\end{align}
which is a set of algebraic equations. Solving \eref{Q_inh} we obtain
\begin{align}
\tilde{Q}_{0}^{inh}(s)=\frac{1}{s+r}\bigg(1+\frac{r}
{s+r+\alpha+\beta}&
[\alpha \tilde{Q}_{1}(s|x_r) \nonumber \\
&+(s+r+\beta)\tilde{Q}_{0}(s|x_r)] \bigg), \nonumber \\
\tilde{Q}_{1}^{inh}(s)=
\frac{1}{s+r}\bigg(1+
\frac{r}
{s+r+\alpha+\beta}&
[\beta\tilde{Q}_{0}(s|x_r) \nonumber \\
&+(s+r+\alpha)\tilde{Q}_{1}(s|x_r)] \bigg) .
\label{Q_inh_sol}
\end{align}
We note that $\tilde{Q}_{\sigma}^{inh}(s)$ depends on $x_r$ and not on $x_0$. Next, we proceed to solve \eref{fpe_lth}. Noting that $x_r$ does not appear in \eref{fpe_lth}, we simply write  $\tilde{Q}_{\sigma}^h(s|x_0)\equiv \tilde{Q}_{\sigma}^h$ for notational convenience. Writing \eref{fpe_lth} in a matrix form, we find
\begin{align}
D\frac{\partial^2}
{\partial x_0^2}
\begin{pmatrix}
\tilde{Q}_{0}^h \\ \\ \tilde{Q}_{1}^h
\end{pmatrix}&
-\lambda \frac{\partial }{\partial x_0} \begin{pmatrix}
\tilde{Q}_{0}^h \\ \\ \tilde{Q}_{1}^h
\end{pmatrix} \nonumber \\
&+\begin{pmatrix}
-(s+\alpha+r)  & \alpha \\ \\ \beta & -(s+\beta+r) 
\end{pmatrix}
\begin{pmatrix}
\tilde{Q}_{0}^h \\ \\ \tilde{Q}_{1}^h
\end{pmatrix}
=0.
\label{fpe_lth_m}
\end{align}
Taking $\Psi=\left(\tilde{Q}_{0}^h\;\;\;\tilde{Q}_{1}^h\right)^T$, we rewrite \eref{fpe_lth_m}
\begin{align}
&D\frac{\partial^2} {\partial x_0^2}\Psi-\lambda \frac{\partial }{\partial x_0}\Psi+\textbf{A} \Psi=\bf{0}, \nonumber \\
&\mbox{where} \;\;\;\;\;\textbf{A}=\begin{pmatrix}-(s+\alpha+r)  & \alpha \\ \\ \beta & -(s+\beta+r) \end{pmatrix}.
\label{fpe_psi}
\end{align}
We now choose $\Psi=\Phi e^{mx_0}$ as the trial solution of \eref{fpe_psi}, that generates the characteristic equation $Dm^2\Phi-\lambda m \Phi+\textbf{A}\Phi=\bf{0}$, which gives
\begin{align}
\begin{pmatrix}Dm^2-\lambda m-(s+\alpha+r)  & \alpha \\ \\ \beta & Dm^2-\lambda m-(s+\beta+r) \end{pmatrix}\Phi=\bf{0}.
\label{fpe_char}
\end{align}
 The roots of \eref{fpe_char} can be found as 
\bea
m_1&=&\frac{\lambda -\sqrt{\lambda ^2+4 D (s+r)}}{2 D},\nonumber\\
m_2&=&\frac{\lambda -\sqrt{\lambda ^2+4 D (\alpha +\beta +s+r)}}{2 D},\nonumber\\
m_3&=&\frac{\lambda +\sqrt{\lambda ^2+4 D (s+r)}}{2 D},
\nonumber\\
m_4&=&\frac{\lambda +\sqrt{\lambda ^2+4 D (\alpha +\beta +s+r)}}{2 D}.
\label{a9}
\eea
A close inspection of the above roots reveals that $m_1$ and $m_2$ are negative, while $m_3$ and $m_4$ are positive (since $D,s,r>0$).
 Since $\tilde{Q}_{\sigma}^h\sim e^{mx_0}$, and $\tilde{Q}_{\sigma}(s|x_0)=1/s$ for $x_0\to\infty$ (because the survival probability $Q_{\sigma}(t|x_0)|_{x_0\to \infty}=1$ and its Laplace transform is $1/s$), we select only $m_1$ and $m_2$ as the plausible roots.\\
\indent
Letting $\Phi_1$ denote the eigenvector corresponding to $m_1$ and utilizing \eref{fpe_char}, we get $\Phi_1=\left(\begin{smallmatrix}1 \\1\end{smallmatrix}\right)$. In a similar manner, the eigenvector corresponding to $m_2$ is given by $\Phi_2=\left(\begin{smallmatrix}-\alpha/\beta \\1\end{smallmatrix}\right)$. The general solution of \eref{fpe_psi} thus reads $\Psi=c_1\Phi_1e^{m_1x_0}+c_2\Phi_2e^{m_2x_0}$, and subsequently
\begin{align}
\tilde{Q}_{0}^h&=c_1e^{m_1x_0}-\frac{\alpha}{\beta}c_2e^{m_2x_0},\label{qh_gsol_1}\\
\tilde{Q}_{1}^h&=c_1e^{m_1x_0}+c_2e^{m_2x_0}
\label{qh_gsol_2}
\end{align}
where $c_1$ and $c_2$ are constants to be determined in the following. To this end, we use the boundary condition $\frac{\partial Q_0(t|x_0)}{\partial x_0}|_{x_0=0}=0$ (equivalently, $\frac{\partial \tilde{Q}_0(s|x_0)}{\partial x_0}|_{x_0=0}=0$) which results in $\frac{\partial \tilde{Q}_0^h(s|x_0)}{\partial x_0}|_{x_0=0}=0$, since $\tilde{Q}_{\sigma}^{inh}(s)$ does not depend on $x_0$.
Applying this in \eref{qh_gsol_1} results in
\bea
c_2=\left(\frac{\beta}{\alpha}\right)\frac{c_1m_1}{m_2}.
\label{c1_c2}
\eea
To compute $c_1$, we recall the other boundary condition (absorbing) $Q_1(t|0)=0$ (equivalently, $\tilde{Q}_1(s|0)=0$).  Combining \eref{Q_inh_sol} and \eref{qh_gsol_2} along with the boundary condition at $x_0=0$, we can write
\bal
\tilde{Q}_{1}(s|0)=c_1+\left(\frac{\beta}{\alpha}\right)\frac{m_1}{m_2}c_1&+\frac{1}{s+r}\bigg(\frac{r}
{s+r+\alpha+\beta}
[\beta\tilde{Q}_{0}(s|x_r) \nonumber \\
&+(s+r+\alpha)\tilde{Q}_{1}(s|x_r)]+1\bigg)
=0.
\label{q0_boun}
\eal
Incorporating \eref{c1_c2} in \eref{q0_boun} and solving for $c_1$ and hence $c_2$ finally gives
\bal
c_1=-\left[\frac{\alpha m_2}{(\alpha m_2+\beta m_1)(s+r)}\right] \bigg(1+&\frac{r}
{s+r+\alpha+\beta}
[\beta\tilde{Q}_{0}(s|x_r) \nonumber \\
&+(s+r+\alpha)\tilde{Q}_{1}(s|x_r)]\bigg),\nonumber\\
c_2=-\left[\frac{\beta m_1}{(\alpha m_2+\beta m_1)(s+r)}\right]\bigg(1+&\frac{r}
{s+r+\alpha+\beta}
[\beta\tilde{Q}_{0}(s|x_r) \nonumber \\
&+(s+r+\alpha)\tilde{Q}_{1}(s|x_r)]\bigg).
\label{c1_c2_sol}
\eal
Plugging in everything together into \eref{Q_h_inh}, we find
\begin{align}
\tilde{Q_0}(s|x_0)=&c_1e^{m_1x_0}-\frac{m_1}{m_2}c_1~e^{m_2x_0}+\frac{1}{s+r}\bigg(1+ \nonumber \\
&\frac{r}
{s+r+\alpha+\beta}
\left[ \alpha \tilde{Q}_{1}(s|x_r)+(s+r+\beta)\tilde{Q}_{0}(s|x_r)\right]\bigg), \nonumber \\
\tilde{Q_1}(s|x_0)=&~c_1~e^{m_1x_0}+\bigg(\frac{\beta}{\alpha}\bigg)\frac{m_1}{m_2}c_1~e^{m_2x_0}+\frac{1}{s+r}\bigg(1+ \nonumber \\
&\frac{r}
{s+r+\alpha+\beta}
\left[ \beta \tilde{Q}_{0}(s|x_r)+(s+r+\alpha)\tilde{Q}_{1}(s|x_r)\right] \bigg),
\label{Q_sol}
\end{align}
which are written in terms of $\tilde{Q}_\sigma(s|x_r)$.  Setting $x_r=x_0$ in \eref{Q_sol} in a self-consistent manner, we find the exact expressions for the survival functions
\bal
&\tilde{Q_0}(s|x_0)=\nonumber\\
&\frac{(\beta m_1+\alpha m_2)+m_1 e^{m_2 x_0} \left(\alpha +\frac{r (\alpha +\beta )}{\alpha +\beta +s}\right)-\alpha m_2 e^{m_1x_0}}{\frac{\beta  m_1 r s e^{m_2 x_0}}{\alpha +\beta +s}+\left(\alpha m_2 r e^{m_1 x_0}+\beta m_1 s+\alpha  m_2 s\right)}, \nonumber \\
&\tilde{Q_1}(s|x_0)=\nonumber\\
&-\frac{(\alpha +\beta +s) \left(\alpha m_2\left(e^{m_1x_0}-1\right)+\beta m_1 \left(e^{m_2 x_0}-1\right)\right)}{(\alpha +\beta +s) \left(\alpha  m_2 r e^{m_1 x_0}+\beta  m_1 s+\alpha  m_2 s\right)+\beta  m_1 r s e^{m_2 x_0}}.
 \label{q_lt2}
\eal
The averaged survival probability can be found by substituting \eref{q_lt2} into \eref{qavg_lt}. This results in
\bal
&\tilde{Q}_{r}^G(s|x_0)=\frac{\beta m_1\left(\frac{r e^{m_2x_0}}{\alpha +\beta +s}+1\right)+\alpha  m_2 \left(1-e^{m_1 x_0}\right)}{\frac{\beta  m_1 r s e^{m_2 x_0}}{\alpha +\beta +s}+\left(\alpha  m_2 r e^{m_1 x_0}+\beta  m_1 s+\alpha  m_2 s\right)}.
 \label{qavg_lt2}
\eal
Generically, \eref{qavg_lt2} can be used to derive all the first passage time moments. The observable of our interest in this problem, for example, the averaged MFPT reads $\langle T_{r}^G\rangle=\tilde{Q}_{r}^G(s|x_0)|_{s=0}$ \cite{gardinar}. This results in \eref{mfpt_av} in the main text.\\
\indent
Alternatively, we can get \eref{mfpt_av} from \eref{Q_sol} by calculating individual mean first passage times conditioned on the initial state of the target state. To this end, let us denote $T_{\sigma}^G(x_0)$ as the first-passage time to reach the target at the origin starting from the position $x_0$ with initial target state being at $\sigma$. Using $\langle T_{\sigma}^G(x_0)\rangle=\tilde{Q_{\sigma}}(s|x_0)|_{s=0}$ and setting $s\to 0$ in \eref{Q_sol}, we get
\begin{eqnarray}
\langle T_{1}^G(x_0)\rangle&=&\frac{1}{r}\left(e^{\mu_1 x_0}-1\right)+
\frac{1}{r}\left[\frac{\beta \mu_1}{\alpha  \mu_2}e^{\mu_1 x_0}\left(1-e^{-\mu_2 x_0}\right)\right], \nonumber \\
\langle T_{0}^G(x_0)\rangle&=&
\frac{1}{r}
\left(e^{\mu_1 x_0}-1\right)\nonumber\\
&+&
\frac{1}{r}\left[\frac{\mu_1}{\mu_2}e^{\mu_1 x_0}\left(e^{-\mu_2 x_0}\left(1+\frac{r}{\alpha}\right)\right. 
+\left.\frac{\beta}{\alpha}\right)\right],
\label{MFPT_sigma}
\end{eqnarray}
where $\mu_1\coloneqq -m_1|_{s=0}>0$ and $\mu_2 \coloneqq -m_2|_{s=0}>0$. Finally, plugging in \eref{MFPT_sigma} into \eref{tavg_def} in the main text, we obtain \eref{mfpt_av}.
\section{The case of confined geometry}
\renewcommand{\theequation}{B.\arabic{equation}}
\setcounter{equation}{0}
\label{appb}
Here we consider the case of a bounded system, i.e., where the particle remains in a finite confinement. This is also highly relevant from the context of chemical reactions, since a high energy barrier can mimic a reflecting boundary (in the reaction coordinate space) that pushes the particle away from it. We construct this finite domain by considering the same set-up as in the main text, with an additional reflecting wall at $x=L>x_0$. If the particle hits the wall it gets reflected back. Intuitively, the barrier or the reflecting boundary prevents the particle from going too far from the target placed at the origin. This is in sharp contrast to the semi-infinite case, where the particle is allowed to diffuse away from the target. This leads to a few key changes in the dynamics that are reflected in the average MFPT.\\
\begin{figure}[t]
    \centering
    \includegraphics[width=8.2cm]{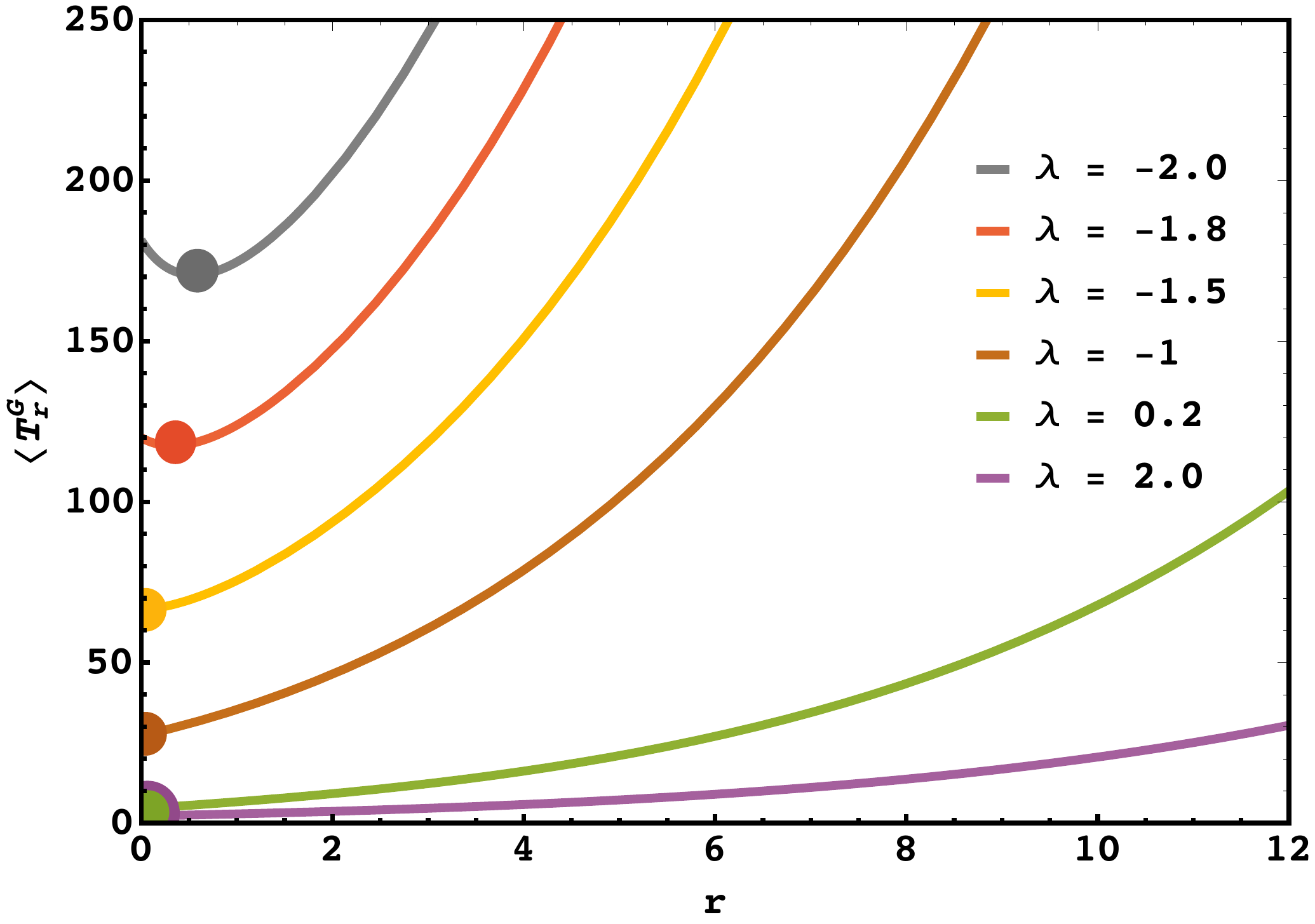}
    \caption{The average MFPT $\langle T_r^G \rangle$ as a function of the resetting rate $r$ for different values of $\lambda$ in the case of bounded domain. Here, the reflecting boundary is placed at $L=3$, where the resetting/initial location is at $x_0=2$. The colored discs mark the optimal resetting rate ($r^{\star}$) in each case. Notably, resetting may not always be helpful ($r^{\star}=0$) even when $\lambda<0$ (e.g., yellow and brown curves).}
    \label{Fig11}
\end{figure}
\indent
On the technical ground, we can find the survival probability in Laplace space using the same method as discussed in \aref{appa} with the new boundary condition $\frac{\partial Q_\sigma(t|x_0)}{\partial x_0}|_{x_0=L}=0$. For this reason, all the eigenvalues will exist unlike in the previous case. In particular, one can show that the eigenvector corresponding to $m_3$ is the same as $m_1$, i.e., $\Phi_1=\left(\begin{smallmatrix}1 \\1\end{smallmatrix}\right)$ and that of $m_4$ is same as $m_2$ i.e. $\Phi_2=\left(\begin{smallmatrix}-\alpha/\beta \\1\end{smallmatrix}\right)$. 
Recalling the decomposition $\tilde{Q}_{\sigma}(s)=\tilde{Q}_{\sigma}^h(s)+\tilde{Q}_{\sigma}^{inh}(s)$ from Eq. (\ref{Q_h_inh}), we first try to obtain the solutions for the homogeneous part.  As before taking $\Psi=\left(\tilde{Q}_{0}^h\;\;\;\tilde{Q}_{1}^h\right)^T$ and using \eref{fpe_psi} we find
\begin{align}
\Psi=c_1\Phi_1e^{m_1x_0}+c_2\Phi_2e^{m_2x_0}+c_3\Phi_1e^{m_3x_0}+c_4\Phi_2e^{m_4x_0},    
\end{align}
so that
\begin{align}
\tilde{Q}_{0}^h&=c_1e^{m_1x_0}-\frac{\alpha}{\beta}c_2e^{m_2x_0}+c_3e^{m_3x_0}-\frac{\alpha}{\beta}c_4e^{m_4x_0},\\
\tilde{Q}_{1}^h&=c_1e^{m_1x_0}+c_2e^{m_2x_0}+c_3e^{m_3x_0}+c_4e^{m_4x_0}.
\end{align}
The inhomogeneous part $\tilde{Q}_{\sigma}^{inh}(s)$ has the same solution as given in \eref{Q_inh_sol}. In addition to the boundary conditions $\tilde{Q}_1(s|0)=0$ and $\frac{\partial \tilde{Q}_0^h(s|x_0)}{\partial x_0}|_{x_0=0}=0$ at the gated target, we also have two additional boundary conditions namely $\frac{\partial \tilde{Q}_0(s|x_0)}{\partial x_0}|_{x_0=L}=0$ and  $\frac{\partial \tilde{Q}_1(s|x_0)}{\partial x_0}|_{x_0=L}=0$. These four boundary conditions give four linear equations  
\begin{align}
    &c_1+c_2+c_3+c_4+\tilde{Q}_{1}^{inh}(s)=0, \nonumber\\
    &c_1m_1e^{m_1L}+c_2m_2e^{m_2L}+c_3m_3e^{m_3L}+c_4m_4e^{m_4L}=0,\nonumber\\
   & c_1m_1e^{m_1L}-\frac{\alpha}{\beta}c_2m_2e^{m_2L}+c_3m_3e^{m_3L}-\frac{\alpha}{\beta}c_4m_4e^{m_4L}=0,\nonumber\\
    &c_1m_1-\frac{\alpha}{\beta}c_2m_2+c_3m_3-\frac{\alpha}{\beta}c_4m_4=0,
    \end{align}
which completely determine the constants $c_1, c_2, c_3, c_4$. 
From this, we can find a closed-form expression for the survival probabilities in Laplace space (and subsequently the MFPT by setting $s\to 0$). The expression for the MFPT is quite lengthy to present here; check \cite{github} for the \textit{Mathematica} file where all the derivations are given. In what follows, we perform a comprehensive analysis of this MFPT and point out the key differences in comparison to the that obtained for the semi-infinite domain.\\
\indent
\fref{Fig11} showcases the variation of the MFPT, $\langle T_r^G \rangle$, as a function of the resetting rate $r$ for different values of the drift $\lambda$. One crucial observation is that the non-monotonic behaviour of $\langle T_r^G \rangle$ is \textit{not always present} even when the drift is away from the target (i.e., $\lambda<0$). This is in complete contrast to the semi-infinite case where resetting is guaranteed to help whenever the drift is away from the target [see Fig. (\ref{Fig3})]. To understand this better, we plot the optimal resetting rate $r^{\star}$ as a function of $\lambda$ for various domain sizes $L$ in \fref{Fig12}, which clearly shows that the critical values of $\lambda$ that marks the resetting transition (denoted $\lambda_c$ in the main text; the minimal value of $\lambda$ for which $r^{\star}=0$), can be negative for considerably smaller domains. With increasing $L$, however, $\lambda_c$ starts to increase and for sufficiently large values of $L$ it saturates to the value of $\lambda_c$ for the semi-infinite case, as displayed \fref{Fig5} of the main text. Simply put, if the reflecting boundary starts moving sufficiently away from the target, the MFPT starts to increase and resetting renders a more effective search in that case. \\
\indent
To elaborate this further, one can expand the MFPT (for finite domain) in the limit of small resetting rate $r\to 0$ and find the first order correction in $r$ [in similar spirit as in \eref{expf} -- see the discussion for the semi-infinite case in Sec. \ref{restart-criterion}, particularly around \eref{separatrix}]. The arguments on the function $f$ as discussed there still holds for the present case (of course, the exact form of the function will be different here). Thus, setting $f=0$ gives us the separatrix distinguishing between the region where resetting helps ($f<0$) and where it hinders ($f>0$). Utilizing this fact, we generate a phase diagram by plotting $\lambda_c$ as a function of $L$, the distance between the reflecting barrier and the origin, and present the same in the inset of \fref{Fig12}. The semi-infinite limit is obtained by taking $L \to \infty$, where $\lambda_c$ saturates to the corresponding value calculated/presented in the main text. For example, we see from \fref{Fig12} that when $p_r=0.5$, $\lambda_c$ saturates to $0.78$, the critical value of $\lambda$ for $p_r=0.5$ (marked in \fref{Fig5} by the vertical yellow line).  \\
\begin{figure}[t]
    \centering
    \includegraphics[width=7.8cm]{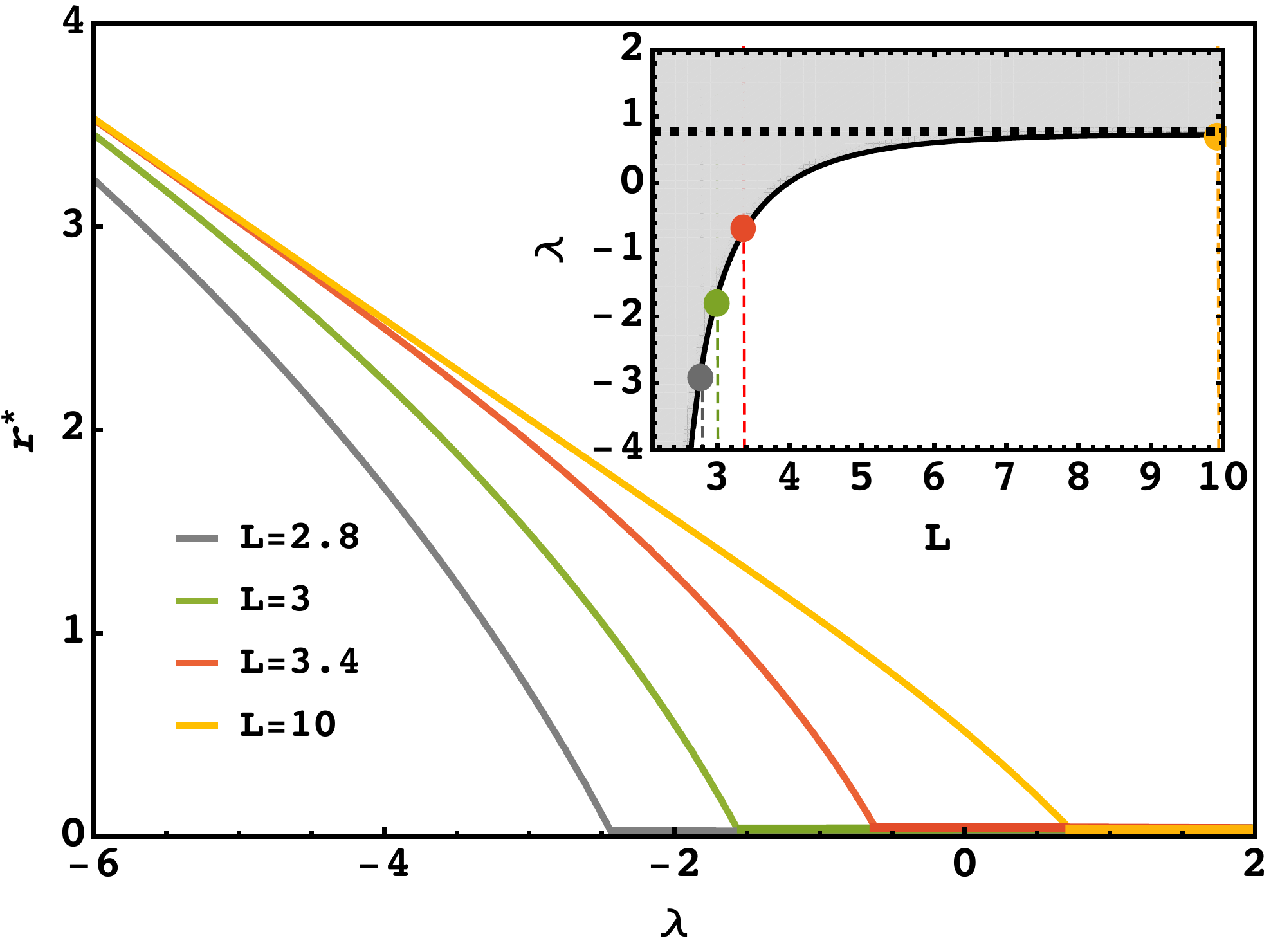}
    \caption{Main: Variation of ORR ($r^{\star}$) as a function of $\lambda$ for different domain size $L$. For $\lambda<\lambda_c$, resetting proves itself beneficial, as indicated by the non-zero values of $r^{\star}$, whereas for $\lambda\ge\lambda_c$, resetting becomes detrimental, as indicated by $r^{\star}=0$. The resetting transition is thus marked by $\lambda_c$. Inset: Phase diagram with reflecting barrier at $L$, by plotting $\lambda_c$ that acts as the separatrix (black line; the colored discs mark the specific cases shown in the main panel) that divides the phase space in two parts; one where resetting is beneficial (white regime) and the other where resetting is detrimental (gray regime). Existence of negative $\lambda_c$ essentially implies that resetting can be detrimental even when $\lambda<0$. The dashed horizontal line corresponds to $\lambda_c=0.78$, obtained for the semi-infinite case (i.e. $L\to\infty$), as displayed in \fref{Fig5} for $p_r=0.5$. Note that we consider $\alpha=0.5,\beta=0.5$ and $D=1$ for each case in the main panel and in the inset.}
    \label{Fig12}
\end{figure}
\vspace{-0.35cm}
\section{The average MFPT for gated drift-diffusion without resetting}
\renewcommand{\theequation}{C.\arabic{equation}}
\setcounter{equation}{0}
\label{appc}
In the absence of resetting, the backward master equations in terms of the survival probability read
\begin{align}
\frac{\partial Q_{0}^{r=0}(t|x_0)}{\partial t}=-&\lambda \frac{\partial Q_{0}^{r=0}(t|x_0)}{\partial x_0}+D\frac{\partial^2 Q_{0}^{r=0}(t|x_0)}{\partial x_0^2}\nonumber\\+&\alpha \left[Q_{1}^{r=0}(t|x_0)
-Q_{0}^{r=0}(t|x_0)\right]\nonumber\\
\frac{\partial Q_{1}^{r=0}(t|x_0)}{\partial t}=-&\lambda \frac{\partial Q_{1}^{r=0}(t|x_0)}{\partial x_0}+D\frac{\partial^2 Q_{1}^{r=0}(t|x_0)}{\partial x_0^2}\nonumber\\+&\beta \left[Q_{0}^{r=0}(t|x_0)
-Q_{1}^{r=0}(t|x_0)\right].
    \label{bfpe-A}
\end{align}
Solving \eref{bfpe-A}, we obtain the following expressions for the survival functions in the Laplace space
\begin{align}
   \tilde{Q_0}^{r=0}(s|x_0)&=\frac{1}{s}-\frac{1}{s} \frac{\alpha  n_2 }{\beta  n_1+\alpha n_2} \left(e^{-n_1 x_0 }-\frac{n_1
   e^{-n_2 x_0 }}{n_2}\right), \nonumber\\
   \tilde{Q_1}^{r=0}(s|x_0)&=\frac{1}{s}-\frac{1}{s} \frac{\alpha  n_2 }{\beta  n_1+\alpha n_2} \left(e^{-n_1 x_0 }+\frac{\beta n_1
   e^{-n_2 x_0 }}{\alpha n_2}\right),
\end{align}
where $n_1=\frac{-\lambda +\sqrt{4 D s+\lambda ^2}}{2 D}$ and $n_2=\frac{-\lambda +\sqrt{4 D (\alpha +\beta +s)+\lambda ^2}}{2 D}$. Setting $s\to 0$, the underlying MFPTs can be computed as before. Eventually, one finds
\begin{align}
    \langle T_0^{r=0} \rangle &=\frac{x_0}{\lambda }+\frac{2 D \left(\beta +\alpha  e^{-\frac{x_0 \left( \sqrt{4
   D (\alpha +\beta )+\lambda ^2}-\lambda\right)}{2 D}}\right)}{\alpha  \lambda 
   \left(-\lambda +\sqrt{4 D (\alpha +\beta )+\lambda ^2}\right)}, \\
    \langle T_1^{r=0} \rangle & =\frac{x_0}{\lambda }+\frac{2 \beta  D \left(-e^{-\frac{x_0 \left( \sqrt{4 D (\alpha +\beta )+\lambda
   ^2}-\lambda\right)}{2 D}}+1\right)}{\alpha  \lambda  \left(-\lambda +\sqrt{4 D (\alpha
   +\beta )+\lambda ^2}\right)}.
\end{align}
The averaged MFPT is then given by 
\begin{align}
 \langle T^G\rangle &= p_r\langle T_{1}^{r=0}\rangle+
 (1-p_r)\langle T_{0}^{r=0}\rangle \nonumber \\
 &=\frac{x_0}{\lambda} +\frac{2 \beta D}{\alpha \lambda \left(\sqrt{4D(\alpha+\beta)+\lambda^2}-\lambda\right)}.
\label{tavg_def-r=0}   
\end{align}
Note that for $p_r\to 1$ ($\beta \to 0$), \eref{tavg_def-r=0} reduces to $\langle T \rangle = x_0/\lambda$, as expected. Moreover, since $D,\alpha,\beta>0$, the second term of the expression of $\langle T^G\rangle$ in \eref{tavg_def-r=0} is always positive, which clearly shows $\langle T^G\rangle> \langle T\rangle$. \\
\vspace{-0.5cm}
\section{Details of numerical simulations}
\renewcommand{\theequation}{D.\arabic{equation}}
\setcounter{equation}{0}
\label{appd}
In this Appendix, we sketch out the basic steps used for the numerical simulations in the main text. We recall that the particle starts from and resets to $x_0$ at a rate $r$. In between reset events, it diffuses in the presence of a drift $\lambda$. Evolution of this particle in microscopic time scale $\Delta t$ can be written in the form of a Langevin dynamics such as 
\begin{eqnarray}
x(t+\Delta t)=
\begin{cases}
 x_0\hspace{3.6cm}\text{w.p. }\;\;\;\;\; r\Delta t \\ \\
 x(t)-\lambda \Delta t +\sqrt{2 D \Delta t}\;\eta(t) \;\;\;\;\text{w.p.}\;\;\;\; (1-r\Delta t) 
\end{cases}
\label{lg}
\end{eqnarray}
where $\eta(t)$ is a $\delta$-correlated white noise i.e., a Gaussian random variable with zero mean and unit variance. Note that the abbreviation w.p. in \eref{lg} has full-form {\it with probability}.  We evolve the particle at each time step $\Delta t$ in our simulation according to \eref{lg} until the particle reaches the target at $x=0$. However, to implement the gating condition to the target, we define a state variable $\sigma=(0,1)$ representing the non-reactive and reactive states of the target, respectively. If target is initially at the non-reactive state, it switches to the reactive state with probability $\alpha \Delta t$. Similarly, it is switched from the active state to the inactive one with probability $\beta \Delta t$. 

To compute the first passage time, we simultaneously track two events: (i) the instant when the particle crosses the origin to the negative side i.e., $x\le 0$ and (ii) note whether the target is in active state i.e., $\sigma=1$. If both the conditions are satisfied, the particle gets absorbed and the process ends. We measure the corresponding time and put it into the first passage statistics. However, if $x\le 0$ and $\sigma=0$ then the particle gets reflected from the boundary, and the process continues till the next absorption occurs. The reflecting boundary condition is implemented by simply reversing the position of the particle i.e. $x \to -x$. The initial target state $\sigma$ is chosen from the steady state i.e.,
\begin{equation}
\sigma(t=0)=
\begin{cases}
1 \hspace{1.0cm} \text{with probability}\;\;\;\;p_r,\\
 0 \hspace{1.0cm} \text{with probability}\;\;\;\; (1-p_r),
\end{cases}
\end{equation}
where $p_r=\alpha/(\alpha+\beta)$. In our simulations, $\Delta t=10^{-5}$ and the averaging was done for $10^5$ successful trajectories for each value of $\lambda$. The results are displayed in \fref{Fig3} with the star symbols, which show excellent agreement with the analytical results presented by the solid lines of same color.

\section*{References:}
 
\end{document}